\documentclass[]{aiaa-tc}
\usepackage[pdftex,pdfauthor={Quindlen},pdftitle={Active Sampling-based Binary Verification of Dynamical Systems}]{hyperref}
\hypersetup{colorlinks,linkcolor={green!50!black},citecolor={green!50!black},urlcolor={blue!80!black}}
\makeatletter \let\NAT@parse\undefined \makeatother

\usepackage{amsmath}
\usepackage{graphicx}
\usepackage{epstopdf}
\usepackage{subfigure}
\usepackage{units}
\usepackage{rotating}
\usepackage{graphicx,paralist}
\usepackage{wrapfig}
\usepackage{nomencl}
\usepackage{amssymb}

\renewcommand{\vec}[1]{\boldsymbol{\mathbf{#1}}} 

\newcommand{\fig}[1] {Figure~\ref{#1}}

\renewcommand{\hat}{\widehat}
\renewcommand{\tilde}{\widetilde}
\newcommand{\vtheta}{\vec{\theta}}

\usepackage{theorem,cite,balance}
\usepackage{algorithm}
\usepackage{algorithmic}
\theoremstyle{plain} \theorembodyfont{\upshape}

\newtheorem{assumption}{\indent Assumption}

\newtheorem{problem}{\indent Problem}

\usepackage[usenames]{color}
\usepackage[svgnames]{xcolor}
\definecolor{DarkGreen}{rgb}{0,0.5,0}
\definecolor{DarkRed}{rgb}{0.75,0,0}
\usepackage{balance}
\usepackage[font=small]{caption}

\usepackage[normalem]{ulem}
\usepackage[capitalize]{cleveref}
\crefformat{equation}{(#2#1#3)}
\Crefformat{equation}{Equation~(#2#1#3)}
\Crefname{equation}{Equation}{Equations}

\renewcommand{\hat}{\widehat}
\renewcommand{\tilde}{\widetilde}

\title{\LARGE \bf
Active Sampling-Based Binary Verification of Dynamical Systems
}

\author{John F. Quindlen$^{1}$%
\thanks{$^{1}$Graduate Student, Department of Aeronautics and Astronautics, Massachusetts Institute of Technology}%
\and Ufuk Topcu$^{2}$\thanks{$^{2}$Assistant Professor, Department of Aerospace Engineering and Engineering Mechanics, University of Texas at Austin}%
\and Girish Chowdhary$^{3}$\thanks{$^{3}$Assistant Professor, Departments of Agricultural \& Biological Engineering and Aerospace Engineering, University of Illinois Urbana-Champaign}%
\and Jonathan P.\ How$^{4}$\thanks{$^{4}$Richard C.\ Maclaurin Professor of Aeronautics and Astronautics, Aerospace Controls Laboratory (Director), MIT}%
}

\begin{document}
\raggedbottom 
\allowdisplaybreaks

\maketitle
\thispagestyle{empty}
\pagestyle{empty}

\begin{abstract}
Nonlinear, adaptive, or otherwise complex control techniques are increasingly relied upon to ensure the safety of systems operating in uncertain environments.  However, the nonlinearity of the resulting closed-loop system complicates verification that the system does in fact satisfy those requirements at all possible operating conditions.  While analytical proof-based techniques and finite abstractions can be used to provably verify the closed-loop system's response at different operating conditions, they often produce conservative approximations due to restrictive assumptions and are difficult to construct in many applications.  In contrast, popular statistical verification techniques relax the restrictions and instead rely upon simulations to construct statistical or probabilistic guarantees.  This work presents a data-driven statistical verification procedure that instead constructs statistical learning models from simulated training data to separate the set of possible perturbations into ``safe'' and ``unsafe'' subsets.  Binary evaluations of closed-loop system requirement satisfaction at various realizations of the uncertainties are obtained through temporal logic robustness metrics, which are then used to construct predictive models of requirement satisfaction over the full set of possible uncertainties.  As the accuracy of these predictive statistical models is inherently coupled to the quality of the training data, an \textit{active} learning algorithm selects additional sample points in order to maximize the expected change in the data-driven model and thus, indirectly, minimize the prediction error.  Various case studies demonstrate the closed-loop verification procedure and highlight improvements in prediction error over both existing analytical and statistical verification techniques.

\end{abstract}

\section{Introduction}

Verification is crucially important for nonlinear and adaptive control systems as they are expected to safely operate in complex, uncertain environments.  In many applications, the control system must ensure the overall system adheres to a wide range of certification criteria.  These criteria include not only closed-loop stability, but other spatial-temporal performance metrics such as aircraft handling and maneuver requirements\cite{MIL1797,FAR25}.  While the entire purpose of the controller is to force the system to satisfy the necessary requirements, the inherent nonlinearity of an adaptive or nonlinear control approach complicates analysis of the closed-loop system's response.  Given slightly different operating conditions, the nonlinearities in the closed-loop dynamics may cause the system to produce drastically different trajectories.  In order to ensure successful operation of the system at different possible operating conditions, verification procedures must demonstrate that the closed-loop dynamical system complies with those certification criteria.  

Various verification procedures exist, but they typically fall within two general categories: analytical proof- or abstraction-based methods and statistical techniques.  If closed-form differential or difference equations of the model are known, then it may be possible to construct analytical certificates that provably verify the stability or boundedness of the closed-loop dynamical system\cite{Moore12_CDC,Prajna05_PhD,Kapinski15_ACC}.  These certificates are based upon analytical functions such as Lyapunov, barrier, or storage functions and separate the bounded set of all possible perturbations into two sets: those from which the system is proven to successfully meet the requirements and those from which the system cannot be proven to meet the requirements.  These analytical certificates are extremely powerful tools for verifying the safety of a control system, but are difficult to implement on many systems.  

One problem with these analytical certificates is that an appropriate certificate is not always obvious nor easy to construct.  Even with recent simulation-guided certificate construction techniques \cite{Kapinski14_HSCC,Topcu08_PhD,Reist15}, it is often difficult to construct a certificate that accurately approximates the true (but unknown) set of all safe operating conditions.  The main limitation is that these certificates are dependent upon the analytical function used during construction.  When the function that best captures the full set of safe perturbations is unknown in advance, then the reliance upon a less-suitable one will result in suboptimal estimates of the true set.  In some applications, such as adaptive controllers, only relatively simple functions are currently available for verification, even though results\cite{Quindlen16_ACC} suggest more complex functions are needed to less-conservatively capture the true set of safe conditions.  

Statistical verification procedures\cite{Clarke11_ATVA,Zhang14_ATVA,Zuliani10_HSCC,Hoxha14_DIFTS,Dang14_ATVA} do not suffer the same limitations as analytical methods since they utilize simulations to directly construct statistical or probabilistic estimates for requirement satisfaction.  While statistical predictions are weaker than analytical guarantees, statistical techniques do not require the same restrictive assumptions and apply to a wider class of systems.  In particular, this work presents data-driven statistical verification techniques\cite{Kozarev16_HSCC} as an alternative to analytical barrier certificates.  Data-driven verification applies machine learning techniques to construct statistical certificates of requirement satisfaction.  At its core, this approach interprets verification as a binary data-classification problem.  Statistical models train on simulations of the closed-loop system and classify potential operating conditions as either ``safe,'' meaning the resulting trajectory will ultimately satisfy the requirement, or ``unsafe,'' meaning it will not.  In this manner, data-driven verification possesses the same applicability of analytical certificates without the reliance upon a particular analytical function.  The removal of this restriction also enables data-driven verification to apply to a much larger class of systems where analytical Lyapunov-like functions are not applicable.  In fact, statistical data-driven methods can be applied to virtually any system with a suitable simulation model or black-box oracle that is able to provide binary safe/unsafe labels for training queries.

While data-driven verification methods are able to relax the conservativeness of analytical certificates and apply to a wider class of problems, the accuracy of their predictions is fundamentally limited by the quality of the simulation data.  Predictions in regions without adequate coverage by training data will have limited accuracy.  
The most straightforward solution to improve the quality of the predictions is simply to saturate the space of possible conditions with a large, structured training dataset using design of experiments techniques\cite{Montgomery13,Lin15}.  While this would help reduce the number of misclassification errors, it neglects the fact simulations can be expensive to obtain.  Especially with a high-fidelity model, each simulation of the closed-loop system may be computationally expensive to obtain, and thus it is impractical and inefficient to blindly saturate the space with a large training dataset.  Instead, it is best to only select the most informative training points.  
The problem is that the ideal training set is unknown until \textit{after} a large set of simulation data has already been obtained.   

This work incorporates active learning\cite{Settles12,Brinker03_ICML,Ali14_AAAI,Kremer14_DMKD,Settles08_EMNLP} into data-driven verification in order to improve the informativeness of the training data without relying upon an exceedingly large dataset.  The expected model change metric\cite{Brinker03_ICML,Kremer14_DMKD} identifies unobserved perturbations which would induce the greatest expected change upon the current statistical classification model.  The resulting sampling algorithms describe two closed-loop verification procedures to maximize the expected informativeness of the training dataset for a designated number of simulations.  These new closed-loop verification procedures directly addresses the limitations of earlier data-driven verification work\cite{Kozarev16_HSCC}, which relied upon passive, open-loop generation of large training datasets.  

The paper is structured as follows. As background, the next two sections provide an overview of the certification problem and discuss the formulation and limitations of analytical verification techniques.  These limitations motivate the data-driven approach towards verification in Section \ref{s:ddverification}.  Section \ref{s:active} presents the closed-loop sampling procedures that iteratively update the data-driven verification model and improves its accuracy.  Subsequent case studies of nonlinear and adaptive control systems demonstrate the improvement in accuracy and sample efficiency.

\section{Problem Formulation}\label{s:prob}

Consider the deterministic nonlinear system
\begin{equation}\label{eq:sys}
	\dot{\vec{x}}(t) = f(\vec{x}(t), \vec{u}(t), \vtheta)
\end{equation}
subject to uncertain operating conditions $\vtheta \in \mathbb{R}^p$, where $\vec{x}(t)\in \mathbb{R}^n$ is the state vector, $\mathcal{X}$ is the set of all feasible states, $\vec{u}(t) \in \mathbb{R}^m$ is the control input vector, and $\mathcal{X}_u$ is the set of all feasible control inputs.  In this problem, the goal is to verify whether a particular control policy is able to ensure the closed-loop system meets all the necessary performance requirements; therefore, the controller which generates inputs $\vec{u}(t)$ is assumed to be known and given a priori.  

The parametric uncertainties $\vtheta$ include all the different operating conditions of the system.  These uncertainties are treated as time-invariant parameters and may arise from a variety of sources.  For example, the perturbation vector $\vtheta$ may include uncertainties about the initial state $\vec{x}(0)$ or system variabilities like vehicle mass and inertia.  
Regardless of their source, the parametric uncertainties $\vtheta$ are considered ``known unknowns'' and are assumed to fall within known bounded set $\Theta$.
\begin{assumption}\label{a:uncountable}
	The set of all possible parametric uncertainties $\vtheta \in \Theta$ is assumed to be a known, compact, and uncountable set $\Theta \subset \mathbb{R}^p$.
\end{assumption}
For instance, aircraft mass and C.G. location will not be perfectly known during flight, but the range of values in which they will lie can be computed in advance.  Thus, verification will perform simulations at various conditions to check whether the aircraft will meet the necessary requirements for all possible mass and loading conditions.  

The trajectory of the system is given by $\Phi(\vec{x}(t)|\vec{x}_0,\vtheta)$, which defines the time evolution of the state vector $\vec{x}(t)$ over the time interval $t \in [0,T_{final}]$.  The nominal initial state vector $\vec{x}_0$ is assumed to be known and fixed while the perturbation vector $\vtheta$ is a particular instantiation from the full set $\Theta$.  As previously mentioned, the actual initial state $\vec{x}(0)$ may vary, but $\vec{x}(0)$ can be modeled as the sum of the fixed, nominal $\vec{x}_0$ and varying $\vtheta$ terms.  Given the deterministic system in \cref{eq:sys}, the closed-loop system's trajectory response is completely determined by these two terms $\vec{x}_0$ and $\vtheta$.  Since $\vec{x}_0$ is fixed, the satisfaction of the performance requirements will only vary due to changes in $\vtheta$.

\subsection{Requirements Modeling}\label{s:requirements}
For simulation-based verification, the performance requirements are assumed to be given ahead of time by relevant certification authorities such as the FAA.  These certification requirements may include relatively straightforward considerations such as stability of the state response, i.e. $\vec{x}(t)$$\rightarrow$$\infty$ as $t$$\rightarrow$$\infty$, and physical bounds on the allowable states such as never-exceed speed of an aircraft.  The criteria might also include more complex spatial-temporal requirements like the avoidance of certain states $\vec{x}(t)$ at particular times.  In order to capture the wide range of possible criteria of interest, it is assumed the requirements can be written with temporal logic specifications.  In short, temporal logic simply provides a mathematical framework for defining requirements and determining whether trajectory $\Phi(\vec{x}(t)|\vec{x}_0,\vtheta)$ satisfies those requirements.  The following material will briefly discuss temporal logic and highlight its application to the simulation-based binary verification problem.

A requirement is specified by a temporal logic formula $\varphi$.  This formula consists of predicate(s) $\zeta$, which are functions of the state and/or control input, as well as boolean and temporal operations on those predicates.  Satisfaction of the temporal logic predicate $\zeta$ is signified by inequality $\zeta>0$.  For instance, if a requirement states that the state $x_1(t)$ must remain below $x_1 = 2$, then the predicate at time $t_i$ is $\zeta[t_i] = 2 - x_1(t_i)$ and the satisfaction of the requirement corresponds to $\zeta[t_i]>0$.  Temporal operations on the predicates are used to encapsulate time dependencies in the requirements.  The three common operators are $\Box_{[t_1,t_2]}$, $\Diamond_{[t_1,t_2]}$, and $\mathcal{U}_{[t_1,t_2]}$, which express that a predicate must hold ``for all'' time within interval $[t_1, t_2]$, ``at some point'' between $t_1$ and $t_2$, and ``until'' another formula is true within $[t_1, t_2]$.  Meanwhile, the boolean operators $\neg$, $\wedge$, and $\vee$ can be used alongside temporal operators to express negation, conjunction, and disjunction.  One of the most important points about temporal logic is that these boolean operators can be used to construct more complex formula from simpler  ones.  For example, formula $\varphi_3 = \Box_{[t_1,t_2]} \varphi_1 \wedge \Diamond_{[t_2,t_3]} \varphi_2$ states that $\varphi_1$ must hold for all times between $t_1$ and $t_2$ \textit{and} $\varphi_2$ must occur at some point between $t_2$ and $t_3$ in order for the formula $\varphi_3$ to be satisfied.  

Once a trajectory $\Phi$ has been generated, a comparison of the state response to the specifications in $\varphi$ determines whether or not the trajectory satisfies the requirement.  Tuple $(\Phi,t) \models \varphi$ signifies that trajectory $\Phi$ satisfies requirement formula $\varphi$ at time $t$, while $(\Phi,t) \models \neg \varphi$ signifies failure to meet the criteria.  This paper uses metric temporal logic (MTL) \cite{Maler04_FORMATS} to produce binary $\{-1,1\}$ measurements of satisfaction from these boolean results.  If the trajectory satisfies the requirement, i.e. $(\Phi,t) \models \varphi$, then measurement $y=1$, else $y=-1$ when the trajectory fails to satisfy the requirement in $\varphi$.  The measurements are rewritten as $y(\vtheta) \in \{-1,1\}$ to emphasize the trajectory and the corresponding satisfaction of the requirement is an explicit function of $\vtheta$ and only changes with $\vtheta$.  While this paper assumes binary measurements, parallel work\cite{Quindlen17a_Arxiv} addresses the case when continuous measurements of satisfaction are available.  These continuous measurements, obtained using signal temporal logic (STL), provide additional information which can be used to approach the problem with a different data-driven verification method.  While this additional information allows for improved quantification of prediction confidence over the methods that will be shown in Section \ref{s:ddverification}, STL is a more restrictive assumption and is not available in all scenarios.  Additionally, active sampling-based verification with binary/discrete measurements can be applied to a much larger class of problems, such as black-box or even human-based labeling of requirement satisfaction.

\subsection{Problem Description}
The goal of the verification procedure is to predict whether a trajectory initialized with the queried $\vtheta$ vector will satisfy the requirements or not.  This binary result naturally leads to the following two sets: set $\Theta_{safe} \subset \Theta$, which contains all $\vtheta$ corresponding to $y(\vtheta) = 1$, and its complement, unsafe set $\Theta_{fail} \subset \Theta$ that contains all $\vtheta$ corresponding to $y(\vtheta) = -1$.  The verification problem can thus be viewed as a binary classification problem, where the goal is to identify whether queried $\vtheta$ vectors belongs to $\Theta_{safe}$ or $\Theta_{fail}$.

\begin{problem}\label{prob:goal}
	The objective is to provide a classifier that separates $\Theta$ into estimates of $\Theta_{safe}$ and $\Theta_{fail}$.  
By definition, $\Theta_{safe}\cup\Theta_{fail}=\Theta$ and $\Theta_{safe}\cap\Theta_{fail}=\emptyset$.  It is assumed $\Theta_{safe}\neq\emptyset$ and $\Theta_{fail}\neq\emptyset$.
\end{problem}

While Assumption \ref{a:uncountable} states the true set $\Theta$ is uncountable, queries can only be made at discrete locations and therefore a simulation-based analysis of $\Theta$ will only ever be able to observe a countable subset of $\Theta$.  With that in mind, it is assumed there is a particular countable approximation of $\Theta$ that should be used for statistical verification.
\begin{assumption}
	There exists a sufficiently-fine discretization of the uncountable set $\Theta$, called $\Theta_d$, which is acceptable for verification purposes.
\end{assumption}
During the data-driven verification process, simulations will be performed at locations $\vtheta\in\Theta_d$.  While the resolution of a sufficiently-fine discretization will change according to the application and possibly the certification authority (such as the FAA), it is assumed the size $|\Theta_d|$ is so large it will be infeasible to perform simulations at every $\vtheta\in\Theta_d$.  To capture this effect, there is assumed to be a cap on the computational budget allocated to data-driven verification.
\begin{assumption}
	The simulation-based data-driven verification procedure is constrained by a computational budget and this budget manifests as a bound on the number of simulations, $N_{total}$, allocated to the verification procedure.
\end{assumption}
If the simulation model used to generate trajectories $\Phi$ is of high fidelity and/or complexity, then the computational overhead required to compute each trajectory will be non-negligible.  Thus, it is unlikely a large number of simulations can be performed within some desired time or budget.  Even when the simulations are relatively cheap, there will still be some feasible upper bound and only a finite number of measurements can be obtained.  Due to the variety of possible causes and constraints, $N_{total}$ is merely intended as a simple constraint to capture the underlying limit on the number of simulations allocated to the verification technique, irrespective of the root cause.

\section{Prior Work: Analytical Barrier Certificates}\label{s:barrier}
The following section will overview prior work\cite{Quindlen16_ACC} in analytical barrier certificates that are the most relevant analytical verification techniques to the problems of interest.  This discussion will serve to not only highlight the strengths of these techniques, but also their limitations.  Two of the examples in Section \ref{s:results} will compare analytical certificates against the results from the paper's data-driven verification procedures.  

\subsection{Approach}
Analytical barrier certificate techniques are a common solution for verification of nonlinear systems.  In particular, these approaches are applied to systems with stability or boundedness requirements, where the state $\vec{x}(t)$ must remain outside of a set of failure states $\mathcal{X}_{fail}$ for the entire trajectory length.  While stability and boundedness covers a wide range of possible problems, it is important to note that this does not include some of the more complex spatial-temporal requirements discussed in Section \ref{s:requirements}.  Thus, these analytical techniques do not apply to all possible systems of interest.  While this does limit their application, analytical barrier certificates are extremely useful and popular techniques whenever appropriate.  

At their core, Lyapunov function-based barrier certificates\cite{Kapinski14_HSCC,Topcu08_PhD,Prajna05_PhD} rely upon a continuously-differentiable scalar function $V(\vec{x}): \mathbb{R}^n \rightarrow \mathbb{R}$ called a Lyapunov function to analyze the performance of the system and bound the trajectory response.  These Lyapunov functions construct invariant sublevel sets that ensure the trajectory never enters $\mathcal{X}_{fail}$.  Given a scalar $\eta>0$ and a Lyapunov function $V(\vec{x})$, an $\eta$-sublevel set $\Omega_{V,\eta}$ defines the set of all $\vec{x} \in \mathcal{X}$ for which $V(\vec{x})$ is bounded below $\eta$: $\Omega_{V,\eta} := \{\vec{x}\in \mathbb{R}^n \ | \ V(\vec{x}) \leq \eta\}$.  
Furthermore, if $V(\vec{x}(0)) \leq \eta$ and $V(\vec{x}(t)) \leq \eta \ \forall t>0$, then this sublevel set is called an invariant sublevel set.  If it can be proven that $V(\vec{x}) > \eta \ \forall \vec{x}\in\mathcal{X}_{fail}$, then any trajectory which starts within the invariant sublevel set $\Omega_{V,\eta}$ will never leave and the trajectory is guaranteed to satisfy the requirement.

\begin{figure}[t]
	\centering
	\includegraphics[width=0.75\columnwidth]{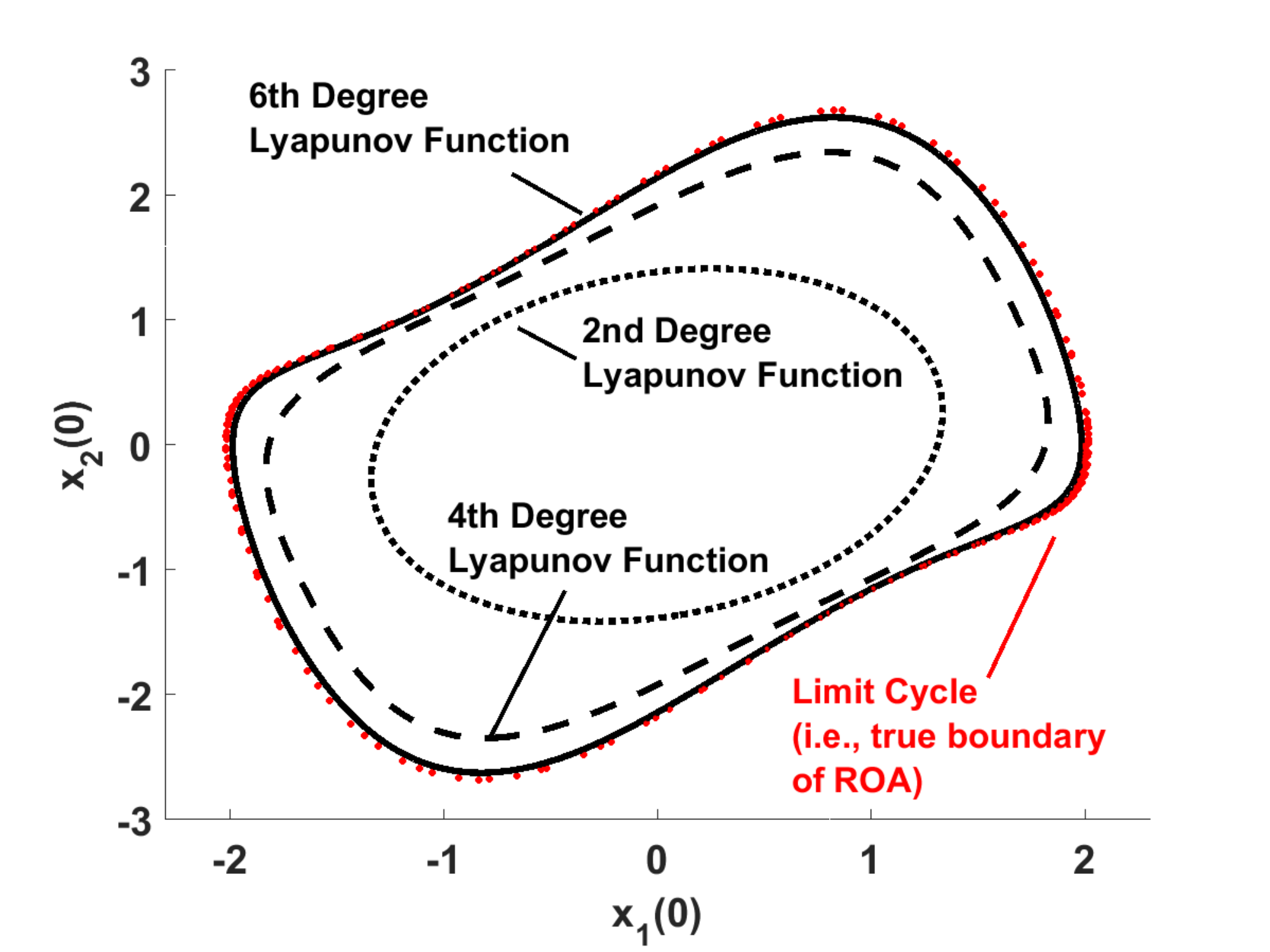}
	\caption{2nd, 4th, and 6th order Lyapunov function-based barrier certificates\cite{Topcu08_Automatica} for the region-of-attraction of an unstable Van der Pol oscillator.  In this problem, the perturbations $\vtheta$ are simply the initial conditions of the second order dynamics, i.e. $\vtheta = [x_1(0), x_2(0)]^T$.  The estimated separation boundaries predicted by the barrier certificates are shown as the black lines, while the true boundary is given by the dotted red line.  The lower order function-based certificates each produce a set of provably-safe $\vtheta$ values, but these sets are conservative and fail to accurately approximate the true boundary.}
 	\label{f:vdp1}
\vspace{-0.1in}
\end{figure}

This invariant sublevel set can then be used to form a barrier certificate to estimate $\Theta_{safe}$.  Given a known $V(\vec{x})$ and $\beta$ that define an invariant sublevel set $\Omega_{V,\beta}$, this invariant sublevel identifies a set of $\vtheta$ values, called $\hat{\Theta}_{V,\beta}$, which are proven to be safe, i.e. $\hat{\Theta}_{V,\beta} \subset \Theta_{safe}$.  The barrier certificate essentially separates $\Theta$ into two classes: those $\vtheta \in \hat{\Theta}_{V,\beta}$ and the remaining $\vtheta$ values which cannot be proven safe with $\Omega_{V,\beta}$.  As they cannot be proven safe, the remaining $\vtheta \not\in \hat{\Theta}_{V,\beta}$ are classified as elements of $\Theta_{fail}$.  
An example of Lyapunov function-based barrier certificates applied to the well-studied unstable Van der Pol oscillator\cite{Topcu08_Automatica} is shown in \fig{f:vdp1}.  In this figure, a 6th order Lyapunov function defines an invariant set (solid black line) which accurately approximates the true separation boundary between $\Theta_{safe}$ and $\Theta_{fail}$ (red dotted line).  $\vtheta$ vectors on or within this $\hat{\Theta}_{V,\beta}$ are proven to be stable (the performance requirement).  

\subsection{Limitations}
While the strength of a theoretically-proven certificate is obvious, these techniques suffer from a variety of limitations which restrict their applicability and help motivate the work in data-driven statistical verification.  As mentioned earlier, not all verification problems can be addressed with analytical certificates.  Even if they do apply, a number of issues may arise which restrict the utility of analytical certificates.  

Lyapunov function-based certificates are conservative by design and can result in vastly overly-conservative predictions with poor construction of the invariant sublevel set $\vtheta \in \hat{\Theta}_{V,\beta}$.  The underlying source of the conservativeness with barrier certificates is that all the remaining $\vtheta \not\in \hat{\Theta}_{V,\beta}$ are not necessarily elements of $\Theta_{fail}$, but because nothing can be proven about these $\vtheta$, they are conservatively classified as elements of $\Theta_{fail}$.  One of the ways to reduce the conservativeness of the barrier certificate is to provide the largest bound $\beta$ for which $\Omega_{V,\beta}$ is still proven safe.  In most cases, this optimal bound is unknown a priori, but simulation-guided techniques\cite{Kapinski14_HSCC,Topcu08_PhD,Reist15} can be used to identify the optimal or near-optimal values.  These procedures take the Lyapunov function $V(\vec{x})$ and generate simulation traces to identify invariant sublevel sets in order to maximize $\beta$.  This simulation-based method was used to construct the certificates found in \fig{f:vdp1}.  

Another source of the conservativeness is the choice of Lyapunov function itself.  If the order or structure of $V(\vec{x})$ is poorly chosen, then even the corresponding optimal $\beta$ will result in a conservative $\hat{\Theta}_{V,\beta}$.  This is seen in the 2nd and 4th order function-based certificates in \fig{f:vdp1}.  These two certificates were constructed with the same simulation-guided method, but were restricted to simpler forms of the Lyapunov function.  Even the simulation-guided $\beta$-maximization technique is handicapped by the incorrect order of the Lyapunov function and cannot accurately predict the true separation boundary.  These results highlight the importance of proper selection of the Lyapunov function as the certificate is highly dependent upon this term.  

The problem is that in many applications, the ``correct'' $V(\vec{x})$ is not immediately obvious or even available.  For instance, if the correct order of a polynomial Lyapunov function is not known, then it may be desirable to start with lower-order functions as the computational complexity is lower and then run simulation-guided techniques\cite{Topcu08_PhD}.  Likewise, higher complexity Lyapunov functions might not be available even with the current state-of-the-art.  For example, model reference adaptive control (MRAC) procedures mostly rely upon quadratic Lyapunov functions to prove stability and convergence\cite{Lavretsky13,Chowdhary10_PhD}.  Recent techniques using these quadratic Lyapunov functions alongside simulation-guided barrier certificate construction\cite{Quindlen16_ACC} were able to obtain a barrier certificate, but this certificate was not able to determine the true $\Theta_{safe}$, as seen later in \fig{f:ex2_init}.  During these situations, it is desirable to apply a supplementary approach not limited by the Lyapunov function $V(\vec{x})$ to explore $\Theta$ and provide feedback as to whether a different Lyapunov function should be used.  

Although the theoretical guarantees provided by analytical certificates are extremely powerful, the issues discussed in this subsection highlight the limitations of even state-of-the-art certificate construction techniques.  The new work presented in the following sections is intended to supplement the existing simulation-guided certificate construction methods or replace them when they aren't applicable.

\section{Data-Driven Verification}\label{s:ddverification}
In place of barrier certificates, data-driven verification will generate statistical classifiers to provide estimates of $\Theta_{safe}$ and $\Theta_{fail}$.  The key insight for data-driven verification is that the verification problem can be re-envisioned as a data classification problem. In particular, since there are only two labels (``safe'' or ``unsafe'') corresponding to satisfaction of the requirements, this is a binary classification problem\cite{Kozarev16_HSCC}.  The resulting statistical classifier must identify whether a particular $\vtheta$ vector belongs to $\Theta_{safe}$ or $\Theta_{fail}$ given a model constructed from a finite training set of observed simulation trajectories.  The following section provides background and describes the initial construction of the statistical certificate during the data-driven verification process.  

\subsection{Initial Training Dataset}
Before anything can be done, the statistical classification model first requires an initial training dataset of simulation trajectories and their binary measurements of requirement satisfaction $y(\vtheta) \in \{-1,1\}$.  Various techniques can be used to generate the initial training points.  The most straightforward approach is to passively select training locations at which the simulations will be performed.  Passive sampling refers to design of experiments (DOE)\cite{Montgomery13,Lin15} or Monte Carlo sampling techniques\cite{Rubinstein04} which do not require any prior information and do not explicitly consider the current statistical model when selecting upcoming training locations.  This black-box, model-agnostic approach makes them appropriate for any type of problem, particularly when there are no available analytical certificates.  The problem with passive sampling procedures is that they may ``waste'' samples in uninformative or well-modeled regions of $\Theta_d$ since samples are taken in all areas of $\Theta_d$ with equal weight or probability.

When data-driven verification augments an existing analytical barrier certificate, the data-driven procedure can exploit this prior information to improve the expected informativeness of the passive training dataset.
If an analytical certificate was already obtained using simulation-guided construction techniques\cite{Topcu08_Automatica}, then the simulation data produced from this method can be used as the statistical classifier's training dataset.  The ``SimLFG'' sampling algorithm\cite{Topcu08_Automatica} used by these analytical techniques randomly selects points along potential sublevel sets of $V(\vec{x})$ as it iteratively searches for an appropriate $\beta$.  The end result is a distribution of sample points, most of which are clustered in close proximity to the boundary of the returned barrier certificate.  In cases where an appropriate $V(\vec{x})$ and $\beta$ are given, but were not constructed using those simulation-guided techniques, the same SimLFG algorithm can still be used by itself to generate the training dataset.  The benefit of generating the samples in this manner is that training samples are concentrated mostly on or outside the current theoretically-proven boundary of the analytical certificate.  This helps avoid the placement of samples within a region of $\vtheta$ values already known to satisfy the requirements and thus ``wasting'' them in uninformative areas.  However, if this information is not available, then the samples must be generated using design of experiments or Monte Carlo sampling.

\subsection{Support Vector Machines As Statistical Certificates}
The data-driven procedure will construct a statistical model to predict the satisfaction of the requirements at queries of $\vtheta$ from the initial training dataset, labeled $\mathcal{L}$, consisting of training locations $\Theta_{\mathcal{L}}$ and corresponding measurements $\vec{y}$.  Given the binary nature of the problem $(y(\vtheta) \in \{-1,1\})$, there are a number of appropriate machine learning/data-mining techniques such as relevance vector machines\cite{Tipping01_JMLR}, kernel logistic regression\cite{Bishop07_PRML}, decision trees\cite{Bombara16_HSCC}, and random forests\cite{Bishop07_PRML}, but the most common and heavily-utilized binary classification approach is support vector machines (SVMs)\cite{Scholkopf02,Vapnik98,Anguita09_DMIN}.  Support vector machines form sets of hyperplanes that separate $\Theta$ into estimates of $\Theta_{safe}$ and $\Theta_{fail}$ when applied to this verification problem.  Additionally, nonlinear support vector machines can efficiently handle arbitrary datasets without modification, even when the training data $\mathcal{L}$ is not linearly separable.  In these cases, the ``kernel trick''\cite{Vapnik98} projects the data into a higher-dimensional representation where the dataset is linearly separable.  Initial work in data-driven verification\cite{Kozarev16_HSCC} has already demonstrated the ability of SVM-based classifiers to accurately separate $\Theta$ in a variety of dynamical systems with linearly- and nonlinearly-separable datasets.

Since it will be unknown in advance if the data is linearly separable, nonlinear SVMs should be used as the default as they will work in both cases.  The nonlinear SVM used for binary classification is given by $H(\vtheta) \in \mathbb{R}$,  
\begin{equation}\label{eq:SVM}
	H(\vtheta) = \sum_{j=1}^{N_{sv}} \alpha_j  y_j \kappa(\vtheta_j,\vtheta) + b,
\end{equation}
where the sign of $H(\vtheta)$ returns a predicted label $\hat{y}(\vtheta)$ for a given query point $\vtheta$,
\begin{equation}\label{eq:prediction}
	\hat{y}(\vtheta) = \text{sign}\big(H(\vtheta)\big).
\end{equation}
During the SVM training process, the procedure selects $N_{sv}$ support vectors from the training set $\mathcal{L}$ and forms the separating hyperplanes used to construct $H(\vtheta)$.  These $N_{sv}\leq|\mathcal{L}|$ support vectors are specifically chosen from the training set to produce the optimal classifier given the current observations.  
Each of these support vectors $\vtheta_j$ has an associated non-zero weight $\alpha_j$, while non-support vector elements of $\mathcal{L}$ can be viewed as having weights $\alpha = 0$.  
The SVM classifier places a nonlinear kernel function $\kappa(\vtheta_j,\vtheta)$ at each support vector location to project the dataset from $\mathbb{R}^p$ to a higher-dimensional space.  Different types of kernel functions are possible, but this work uses the common isotropic radial basis function (RBF) kernel.  This isotropic RBF kernel function is given by  
\begin{equation}\label{eq:rbf}
	\kappa(\vtheta_j,\vtheta) = \text{exp}\Big(\frac{-||\vtheta_j - \vtheta||^2}{\gamma^2}\Big),
\end{equation}
where $\vtheta_j$ is the location of the indicated support vector and scalar hyperparameter $\gamma$ is the kernel width.  

In order to compute the optimal set of support vectors from $\mathcal{L}$ and their associated weights $\alpha$, the support vector machine training algorithm solves the following optimization program.  Equation~\ref{eq:training} gives the Lagrangian dual of the primal quadratic program for soft-margin nonlinear SVMs\cite{Scholkopf02},
\begin{align}\label{eq:training}
	\underset{\vec{\alpha}}{\text{maximize}} & \sum_{j=1}^{N_{sv}} \alpha_j - \frac{1}{2} \sum_{j=1}^{N_{sv}}\sum_{i=1}^{N_{sv}} \alpha_i \alpha_j y_i y_j \kappa(\vtheta_j, \vtheta_i) \\
		\text{s. t.} & \sum_{j=1}^{N_{sv}} \alpha_j y_j = 0 \\
		& 0 \leq \alpha_j \leq C_{\alpha}. \label{eq:training2}
\end{align}
Rather than compute the solution using the primal form, it is generally easier to solve the problem in this dual form.  
Before the training process can begin, two hyperparameters must be set by the programmer: the $\alpha$ box constraint $C_{\alpha}$ from \cref{eq:training2} and the kernel width $\gamma$ from \cref{eq:rbf}.  The choice of hyperparameters will affect the selection and number of support vectors.  In this paper, the kernel width is kept fixed to $\gamma = 1$, although it can also be estimated at part of a hyperparameter optimization step within the training process.  Equation \ref{eq:SVM} also includes an optional bias term $b$, but this bias is set to $b = 0$ hereafter.  The end result of this training procedure is the statistical classifier shown in \cref{eq:SVM}.  The classifier and the resulting predictions $\hat{y}(\vtheta)$ for all $\vtheta \Theta_d$ will then define two sets, $\hat{\Theta}_{safe}$ and $\hat{\Theta}_{fail}$.


\subsection{Prediction Error}\label{s:errors}
As with analytical certificates, data-driven certificates are susceptible to misclassifications when labeling unseen locations in $\Theta$.  Like before in Section \ref{s:barrier}, some points that were labeled as ``unsafe'' by the classifier, $\vtheta \in \hat{\Theta}_{fail}$, may actually belong to $\Theta_{safe}$; however, data-driven classifiers also introduce the dangerous possibility that unobserved unsafe locations $\vtheta \in \Theta_{fail}$ will be incorrectly labeled as ``safe'' and included in $\hat{\Theta}_{safe}$.  While the possibility of these errors cannot be completely eliminated, a number of steps can be taken during training to estimate their likelihood and minimize their impact.  

First, both types of misclassification errors have unequal consequences.  In general, it is much more acceptable to accidentally label $\vtheta \in \Theta_{safe}$ as ``unsafe'' than the reverse.  The scalar constraint term $C_{\alpha}$ from the training process can be recast as a matrix to reflect the unequal cost,
\begin{equation}
	C_{\alpha} = \begin{bmatrix} C_{FN} & 0 \\ 0 & C_{FP} \end{bmatrix}.
\end{equation}
Higher $C_{FP}$ values will place higher training penalties on false positives, that is incorrect labeling of unsafe points as ``safe.''   Meanwhile, lower $C_{FN}$ will loosen the penalties on the false negatives, when safe points are incorrectly labeled as ``unsafe.''  In this manner, the optimization program will adjust the support vectors and their weights to avoid false positives, potentially at the cost of more false negatives.  

Additionally, while the $C_{\alpha}$ constraint will help control the number of misclassification errors, it is impossible to completely eliminate the possibility they exist.  Therefore, it is advantageous to estimate the likelihood of misclassification errors, particularly in unobserved spaces of $\Theta$.  The primary method to estimate the likelihood of misclassification erros is to compute the numerical generalization error on an independent validation dataset.  This independent validation dataset contains observed datapoints and their corresponding measurements where the accuracy of the classifier's predictions can be tested.  The most common approaches are leave-one-out and k-fold cross-validation approaches which segment the training data $\mathcal{L}$ into a set of training points actually used to construct $H(\vtheta)$ and an independent testing set used to validate the performance of $H(\vtheta)$\cite{Anguita09_DMIN} on ``held-out'' data.  

These two validation approaches can be used to estimate the likelihood of misclassification error, but they are limited in some aspects.  First, the accuracy of the cross-validation error with respect to the true generalization error over $\Theta$ is limited by the quality of $\mathcal{L}$.  If the observed training dataset does not adequately cover a region in $\Theta$, then the independent testing set will fail to accurately incorporate its effect upon the generalization error.  Likewise, the creation of independent, held-out datasets from the training set, reduces the amount of already-limited training data that can be used to construct the classifier.  Simply put, all the data held out from $\mathcal{L}$ to create the independent validation set, could have been used to produce a better classifier.

While their presence is noted, these limitations are an unfortunate effect of the reliance upon binary measurements.  As mentioned in Section \ref{s:prob}-\ref{s:requirements}, parallel work\cite{Quindlen17a_Arxiv} has developed a similar data-driven approach when continuous-valued measurements of requirement satisfaction are available.  In that work, prediction confidence is computed online without having to segment $\mathcal{L}$ due to the extra layer of information provided by continuous measurements.  While this addresses most of the misclassification issues discussed above, continuous-valued signal temporal logic (STL) measurements are not always available.  For systems with binary MTL measurements, STL-like prediction confidence can be estimated without a separate validation set using Platt scaling\cite{Platt99_ALMC}.  This method fits a logistic regression model to the SVM classifier and produces probabilistic estimates of prediction confidence much like systems with continuous STL measurements\cite{Quindlen17a_Arxiv}.  Unfortunately, as with kernel logistic regression, Platt scaling assumes linearly separable data, which limits its accuracy when applied to nonlinearly-separable datasets.  Regardless of their respective drawbacks, many methods exist for the computation of prediction error in the data-driven verification process.  

\section{Closed-Loop Verification}\label{s:active}
Section \ref{s:ddverification} described a method for constructing the statistical certificate given an initial set of training data.  As with all statistical learning approaches, data-driven verification is fundamentally limited by the training dataset used to construct the certificate.  If the data is limited to only a small region of $\Theta$, then the statistical certificate's accuracy will degrade past this small region.  Previous work in data-driven verification\cite{Kozarev16_HSCC} successfully sidestepped this issue by generating very large training datasets of randomly-selected points that implicitly covered $\Theta$ with adequate resolution.  These large training sets pose a problem if simulation trajectories are computationally expensive to obtain or the number of simulations budgeted for the verification procedure is small.  In an ideal scenario, simulations would only be performed at the most-informative $\vtheta$ locations distributed along the true boundary (or boundaries) that separates sets $\Theta_{safe}$ and $\Theta_{fail}$.  This training set would minimize the predictive error without relying upon a large number of simulation samples; however, such a training dataset would require the true boundary to be known in advance.  

The work in this section exploits active learning to cluster samples near the estimated boundary and approximate the ideal training set without any prior knowledge of the true boundary.  The main contribution of this work is a novel verification process that iteratively selects potentially-informative training points in order to maximize the accuracy of the predictions.  Due to the iterative nature of the data-driven verification process, we call the approach \textit{closed-loop verification}.  The following subsections will present methods to identify informative locations for future simulations as well as detail two closed-loop verification procedures.  

\subsection{Selection Criteria}
Active learning describes a closed-loop sampling procedure that iteratively trains a statistical model and selects new sample points to improve the current objective\cite{Settles12,Brinker03_ICML,Ali14_AAAI,Kremer14_DMKD,Settles08_EMNLP}. Rather than passively select training locations, either in a grid or random fashion, active learning uses the current model information to discriminate between potential sample locations and select new training locations which are expected to best improve the current model.
Depending on the objective, the ``best'' possible next sample points might be different and there are a number of different sample selection criteria corresponding to various objectives.  A comprehensive overview of some of these different objectives and selection criteria is found in Settles' book\cite{Settles12}.  At its core, active learning is a closed-loop process that exploits the most recent information, regardless of the particular selection criteria.  This allows the sampling procedure to minimize the expected prediction error of the classifier given a designed number of allowable simulations, and thus avoiding the reliance upon large datasets used in the earlier open-loop data-driven verification procedures\cite{Kozarev16_HSCC}.  

\subsubsection{Expected Model Change}
Due to the desire to minimize the rate of misclassification errors, this work uses the expected model change metric\cite{Kremer14_DMKD} to rank the prospective sample locations and identify which location is expected to most improve the current data-driven verification model.  
The objective is to select the sample point that, if measured, would induce the largest expected change upon the current model once the model is retrained with this new information.  
For SVMs, it was shown\cite{Kremer14_DMKD} that the point which maximizes expected model change can be found through the gradient of the Lagrangian dual objective function in \cref{eq:training}.  If a new sample location, labeled $\vtheta_{+1}$, and its hypothetical measurement $y(\vtheta_{+1})$ are added to the current SVM, the resulting dual objective would be given by
\begin{equation}\label{eq:tempObj}
	D(\vec{\alpha}) = \sum_{j=1}^{N_{sv}+1} \alpha_j - \frac{1}{2} \sum_{j=1}^{N_{sv}+1}\sum_{i=1}^{N_{sv}+1} \alpha_i \alpha_j y_i y_j \kappa(\vtheta_j, \vtheta_i)\ .
\end{equation}
With the weighting term $\alpha_{+1}$ initialized to zero, the gradient of \cref{eq:tempObj} with respect to $\alpha_{+1}$ is then
\begin{equation}\label{eq:tempGrad}
	\frac{\partial D(\vec{\alpha})}{\partial \alpha_{+1}} = 1 - y(\vtheta_{+1}) \sum_{j=1}^{N_{sv}} \alpha_j  y_j \kappa(\vtheta_j,\vtheta_{+1})\ .
\end{equation}
Assuming the bias $b$ from \cref{eq:SVM} is set to 0, this gradient is simply
\begin{equation}\label{eq:tempGrad2}
	\frac{\partial D(\vec{\alpha})}{\partial \alpha_{+1}} = 1 - y(\vtheta_{+1})H(\vtheta_{+1})\ ,
\end{equation}
where $H(\vtheta_{+1})$ is the output of the current SVM model at location $\vtheta_{+1}$.  Since $\alpha_{+1}$ must be non-negative, the model should only be updated if \cref{eq:tempGrad2} is positive, meaning $y(\vtheta_{+1})H(\vtheta_{+1}) < 1$.  Note that in order for $y(\vtheta_{+1})H(\vtheta_{+1}) < 0$, there must be disagreement between the sign of current prediction $H(\vtheta_{+1})$ and the actual measurement $y(\vtheta_{+1})$.  If this actual $y(\vtheta_+1)$ was indeed known, then the optimal sample $\overline{\vtheta}$ which maximizes the model change would induce the largest gradient in \cref{eq:tempGrad2},
\begin{equation}\label{eq:infeasibleUnc}
	\overline{\vtheta} = \underset{\vtheta_{+1} \in \Theta}{\text{argmax }} \Big(1 - y(\vtheta_{+1})H(\vtheta_{+1})\Big).
\end{equation}
The problem with the selection metric in \cref{eq:infeasibleUnc} is that $y(\vtheta_{+1})$ is not known until after a simulation has already been performed at that location.  In place of the actual model change, the \textit{expected} model change is used as the selection metric.  This expected model change is given by
\begin{equation}\label{eq:feasibleUnc}
	\overline{\vtheta} = \underset{\vtheta_{+1} \in \Theta}{\text{argmax }} \Big(1 - \hat{y}(\vtheta_{+1}) H(\vtheta_{+1})\Big),
\end{equation}
with $\hat{y}(\vtheta_{+1}) = \mathbb{E}[y(\vtheta_{+1})]$ as the expected measurement at $\vtheta_{+1}$.  Shown earlier in \cref{eq:prediction}, the expected measurement $\hat{y}(\vtheta_{+1})$ is actually the sign of the output of the current classifier $H(\vtheta_{+1})$, forcing $\hat{y}(\vtheta_{+1}) H(\vtheta_{+1}) > 0$ to always hold true.  Therefore, the ``best'' sample point according to the expected model change \cref{eq:feasibleUnc} is simply the point whose classifier output $H(\vtheta)$ is closest to zero, i.e. $|H(\overline{\vtheta})| \approx 0$, since $\hat{y}(\vtheta_{+1}) H(\vtheta_{+1})$ will never be negative.  This premise forms the basis of the active sampling procedure used for closed-loop verification.

\subsection{Sequential Sampling}
\cref{alg:seqSampling} details the sequential closed-loop verification procedure.  The procedure selects samples based upon the aforementioned expected model change metric, rewritten as
\begin{equation}\label{eq:singleUnc}
	\overline{\vtheta} = \underset{\vtheta \in \mathcal{U}}{\text{argmin }} |H(\vtheta)|.
\end{equation}
The procedure assumes samples are taken from a set of unobserved samples locations, set $\mathcal{U}$, such that $\Theta_{\mathcal{L}}\cap\mathcal{U} = \emptyset$.  
Given the existence of the sufficiently-fine $\Theta_d$ and an initial training set $\mathcal{L}$ taken from $\Theta_d$, the unobserved set $\mathcal{U}$ is simply the remainder $\mathcal{U} = \Theta_d\setminus \Theta_{\mathcal{L}}$.  Additionally, it is assumed the discretization $\Theta_d$ is so fine that the final size of $\mathcal{U}$ is still significantly larger than $N_{total}$, i.e. $|\mathcal{U}| \gg N_{total}$.  

The closed-loop process is as follows.  Given the initial training set $\mathcal{L}$ and classification model $H(\vtheta)$, the procedure in \cref{alg:seqSampling} selects the remaining $T = N_{total} - |\mathcal{L}|$ samples from $\mathcal{U}$. Once the next training location $\overline{\vtheta}$ has been selected (line 3), a simulation is performed using that $\vtheta$ vector and the binary MTL measurement is obtained (line 4).  The new measurement is added to the training set $\mathcal{L}$ (line 5) and the model is retrained with this new dataset (line 6).  This process repeats until the size of $\mathcal{L}$ reaches $N_{total}$.  
\begin{algorithm}[!]
\caption{Sequential active learning using the expected model change selection metric}
\label{alg:seqSampling}
\begin{algorithmic}[1]
	\STATE \textbf{Input:} training set $\mathcal{L}$, available sample locations $\mathcal{U}$, initial classifier $H(\vtheta)$, \# of additional samples $T$
	\FOR{i=1:T}
	\STATE{Select $\overline{\vtheta}$ from $\mathcal{U}$ according to \cref{eq:singleUnc}}
	\STATE{Run simulation at $\overline{\vtheta}$, obtain measurement $y(\overline{\vtheta})$}
	\STATE{Add $\{\overline{\vtheta},y(\overline{\vtheta})\}$ to training set $\mathcal{L}$, remove $\overline{\vtheta}$ from $\mathcal{U}$}
	\STATE{Retrain $H(\vtheta)$ with new $\mathcal{L}$}
	\ENDFOR
\end{algorithmic}
\end{algorithm}

\subsection{Batch Sampling}
The procedure described in \cref{alg:seqSampling} is a sequential process - it iteratively updates the model one point at a time.  While this process will correctly guide the selection of datapoints as intended, it ignores two important considerations.  First, sequential sampling procedures do nothing to address the nontrivial cost of retraining the classifier.  As the number of additional training points in $\mathcal{L}$ grows, so does the computational effort required to recompute classifier model.  If the model is retrained after each the arrival of every single measurement, then the computational cost associated just with retraining in high.  Additionally, sequential sampling algorithms do not exploit the parallelism inherent in many simulation environments.  In particular, multiple simulations can often be performed in parallel due to multiple processor cores or computers.  Batch active learning processes\cite{Kremer14_DMKD} address both these considerations by selecting multiple $\overline{\vtheta}$ values before retraining $H(\vtheta)$.  If the number of remaining samples $T$ is broken into $T_B$ batches of $M$ points, the computational cost of retraining decreases from $\mathcal{O}(N^2T)$ to $\mathcal{O}(N^2T_B/M)$.  

While the selection of a batch of points between retraining steps will reduce the cost associated with retraining the SVM, it introduces another concern.  If the same criteria from \cref{eq:singleUnc} is blindly implemented without care, the fine resolution of points in $\mathcal{U}$ can result in the selection of multiple neighboring points in close proximity as they will have similar values according to \cref{eq:singleUnc}.  In order to prevent the selection of redundant points, a diversity measure must be incorporated into the selection criteria.  This paper uses the angle between the hyperplanes $\phi(\vtheta)$ induced by the samples as the diversity metric.  The angle associated with a kernel $\kappa$ is shown in \cref{eq:diversity}.  From a diversity perspective, the goal is to maximize the angle between samples.
\begin{equation}\label{eq:diversity}
	|\cos(\angle(\phi(\vtheta_i),\phi(\vtheta_j)))| = \frac{|\kappa(\vtheta_i,\vtheta_j)|}{\sqrt{\kappa(\vtheta_i,\vtheta_i)\kappa(\vtheta_j,\vtheta_j)}}
\end{equation}
The batch selection criteria metric modifies the sequential sample selection criteria in \cref{eq:singleUnc} with the diversity measure from \cref{eq:diversity}.  This new selection metric becomes
\begin{equation}\label{eq:batchUnc}
	\overline{\vtheta} = \underset{\vtheta \in \mathcal{U}}{\text{argmin }}\big( \lambda |H(\vtheta)| + (1-\lambda) \underset{\vtheta_j \in \mathcal{S}}{\text{max }}|\cos(\angle(\phi(\vtheta),\phi(\vtheta_j)))|\big)\ ,
\end{equation}
which is a weighted combination of the expected model change and diversity in the batch.  Here, the batch sampling metric selects the $M$ points in the batch sequentially and set $\mathcal{S}\in\Theta$ refers to the points previously selected for the current batch. The scalar term $\lambda \in [0,1)$ controls the weighting between the metrics \cref{eq:singleUnc} and \cref{eq:diversity}.  As $\lambda$ grows, the diversity metric has less effect upon $\overline{\vtheta}$ and the selection criteria converges towards \cref{eq:singleUnc}.  This work uses $\lambda = 0.7$ as it has demonstrated good empirical results. 
This balance of utility and diversity ensures that each subsequent sample chosen by \cref{eq:batchUnc} is adequately different from all the preceding points in $\mathcal{S}$.  

The complete batch active learning algorithm is given in \cref{alg:uncSampling}.  On line 1, the procedure starts with the initial training set $\mathcal{L}$ and classifier $H(\vtheta)$ along with the set of available sample points $\mathcal{U}$.  The algorithm the iteratively selects the ``best'' data point according to \cref{eq:batchUnc} (line 5) and adds this $\overline{\vtheta}$ to set $\mathcal{S}$ (line 6).  Once $\mathcal{S}$ is full, the simulations are performed at these $\vtheta$ locations (line 9).  On line 10, set $\mathcal{S}$ and the corresponding labels are added to the training set $\mathcal{L}$ and this new information is used to retrain $H(\vtheta)$.  The closed-loop verification process repeats lines 3-11 until a designated number of iterations has been reached.  As will be illustrated in Section \ref{s:results}, Algorithm \ref{alg:uncSampling} successfully distributes the sample locations across regions of high utility (large expected model change) and implicitly improves the prediction error over passive sampling procedures.  

\begin{algorithm}[t]
\caption{Batch active learning using the expected model change selection metric}
\label{alg:uncSampling}
\begin{algorithmic}[1]
	\STATE \textbf{Input:} unlabeled set  $\mathcal{U}$, training set $\mathcal{L}$, batch size $M$, classifier $H(\vtheta)$, empty set $\mathcal{S}$
	\FOR{each iteration}
	\STATE \textbf{Reset:} $d=1$
	\WHILE{$d \leq M$}
	\STATE{Select $\overline{\vtheta}$ from $\mathcal{U}$ according to \cref{eq:batchUnc}}
	\STATE{Add $\overline{\vtheta}$ to $\mathcal{S}$, remove $\overline{\vtheta}$ from $\mathcal{U}$}
	\STATE{$d = d+1$}
	\ENDWHILE
	\STATE{Run simulations $\forall \vtheta\in\mathcal{S}$, label the results}
	\STATE{Add set $\mathcal{S}$ to training set $\mathcal{L}$, reinitialize empty $\mathcal{S}$}
	\STATE{Retrain $H(\vtheta)$ with new $\mathcal{L}$}
	\ENDFOR
\end{algorithmic}
\end{algorithm}

\begin{figure}[!]
	\centering
	\includegraphics[width=0.6\columnwidth]{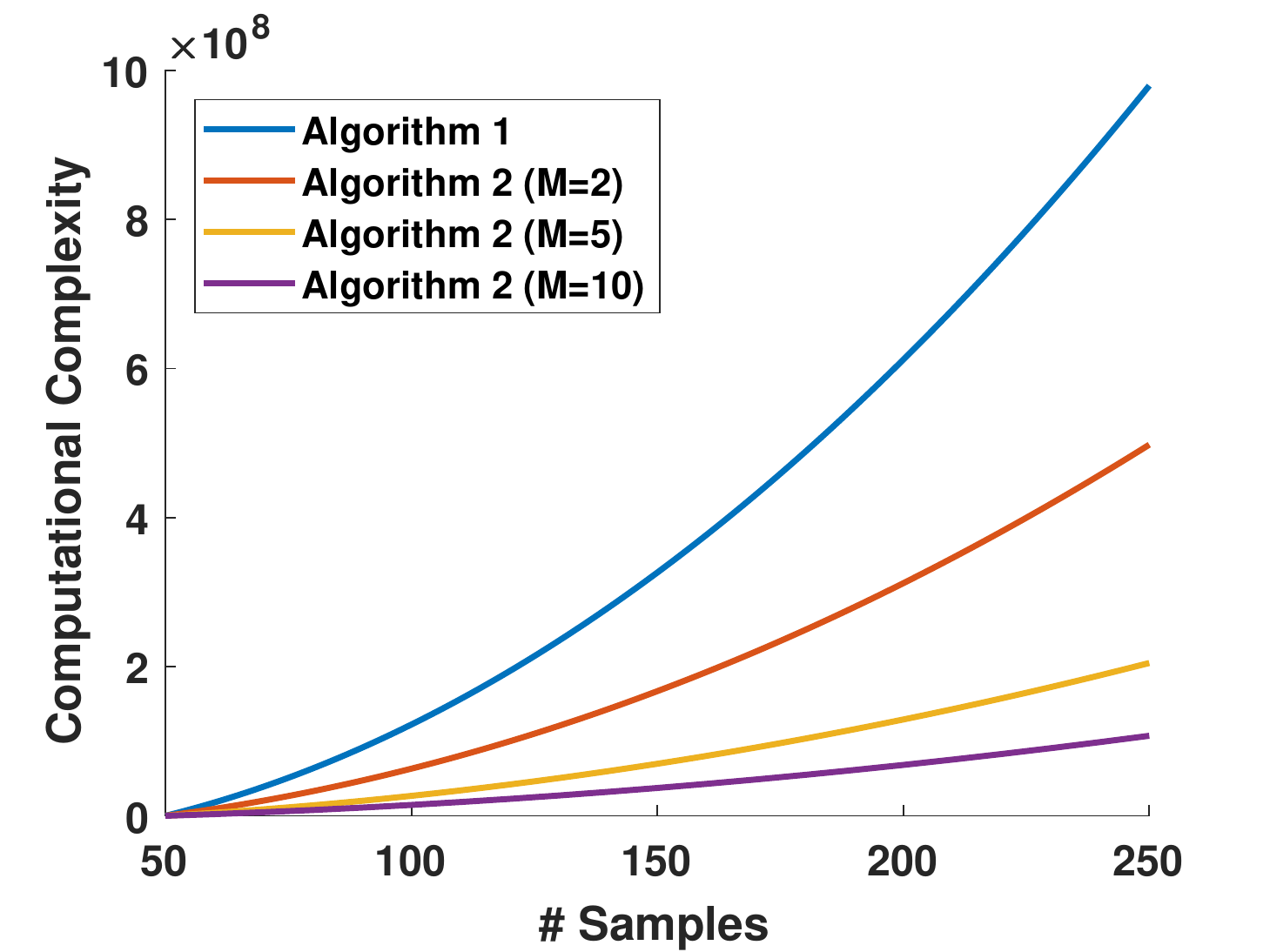}
\vspace{-0.1in}
	\caption{Computational complexity of the sequential (Algorithm \ref{alg:seqSampling}) and batch (Algorithm \ref{alg:uncSampling}) closed-loop verification procedures when applied to Example \ref{s:results}-\ref{s:clmrac_ex1}.  By reducing the number of retraining steps for the sample number of total samples, Algorithm \ref{alg:uncSampling} lowers the computational cost required for closed-loop verification.  The exact reduction varies according the particular example, but a larger batch size $M$ will lower the complexity.}
	\label{f:clmrac_cost}
\end{figure}

Although Algorithm \ref{alg:uncSampling} requires more intermediate steps between retraining the SVM, the net computation cost is lower than in Algorithm \ref{alg:seqSampling}.  As mentioned in the preceding paragraphs, the training step in Algorithm \ref{alg:seqSampling} (line 6) and Algorithm \ref{alg:uncSampling} (line 11) requires $\mathcal{O}(N^2)$ operations for each iteration.  In Algorithm \ref{alg:seqSampling}, the prediction step (line 3) requires $\mathcal{O}(N|\mathcal{U}|)$ operations to weight the prospective sample locations.  Note that this does not include the cost of actually performing simulations, as that will vary from example to example.  For comparison, the batch sample selection step in line 5 of Algorithm \ref{alg:uncSampling} requires an additional $\mathcal{O}(M|\mathcal{U}|)$ operations to calculate the diversity measure for all $M$ points in the batch.  Although this requires additional operations, the total computational cost for the batch procedure is lower for an equal number of samples.  Figure \ref{f:clmrac_cost} demonstrates the computational complexity for the two procedures when applied to Example \ref{s:results}-\ref{s:clmrac_ex1}.  Algorithm \ref{alg:uncSampling} produces a lower complexity than the sequential procedure and this improvement in computational complexity increases with larger batch sizes $M$.

\section{Results}\label{s:results}
The closed-loop data-driven verification procedure is demonstrated on a number of case studies.  The results for each example are summarized, but the first two case studies also have a link to accompanying videos and supplementary material.  These videos best illustrate the iterative nature of the approach.  Additionally, the first example highlights that data-driven verification can be applied to a wider array of problems, not just adaptive control problems.  The third example also demonstrates data-driven analysis on a verification problem which cannot be addressed with current analytical barrier certificate techniques.

\subsection{Unstable Van der Pol Oscillator}
The first example is not an adaptive control system, but rather the well-studied Van der Pol oscillator problem shown previously in \fig{f:vdp1}.  This example not only demonstrates the use of data-driven verification to supplement existing analytical barrier certificates, but also highlights its application to a wider class of systems than adaptive systems.  The nonlinear dynamics in \cref{eq:vdp} have an unstable limit cycle and an asymptotically stable equilibrium point at the origin,
\begin{align}\label{eq:vdp}
	\dot{x}_1 & = -x_2 \\
	\dot{x}_2 & = x_1 + (x_2^2 -1)x_2. \notag
\end{align}
\noindent The verification goal is to identify initial conditions $\vtheta = [x_1(0) \ x_2(0)]^T$ that will cause the system to either converge to the origin or diverge.  The region of $\vtheta$ values for which the trajectory converges, $\Theta_{safe}$, is also known as the region-of-attraction (ROA).

\begin{figure}[t]
	\centering
	\subfigure[Initial $\mathcal{L}$ and the barrier certificate produced with a quadratic Lyapunov function]{\includegraphics[width=0.49\columnwidth]{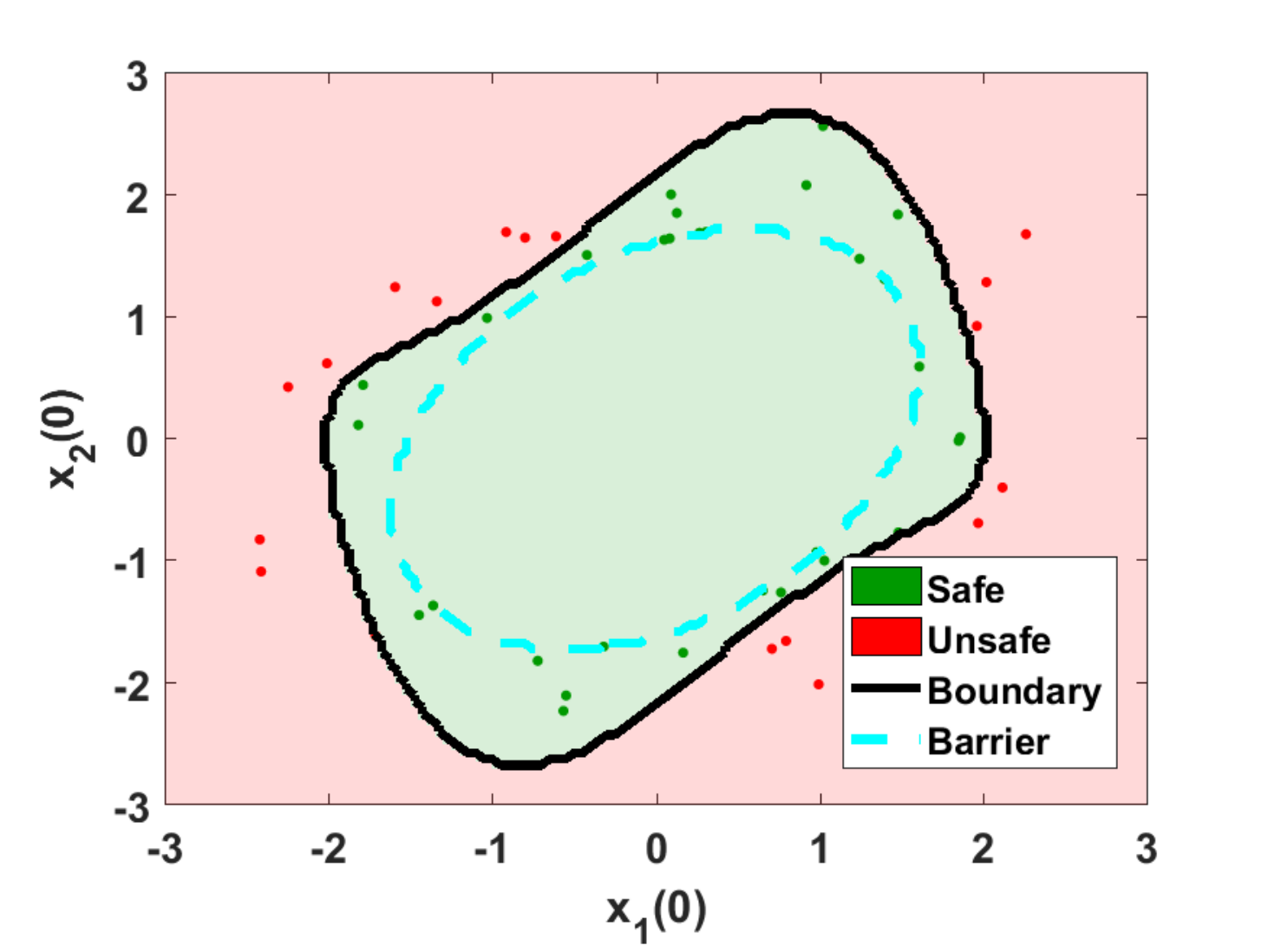}}
	\subfigure[Starting SVM model trained using the initial training set]{\includegraphics[width=0.49\columnwidth]{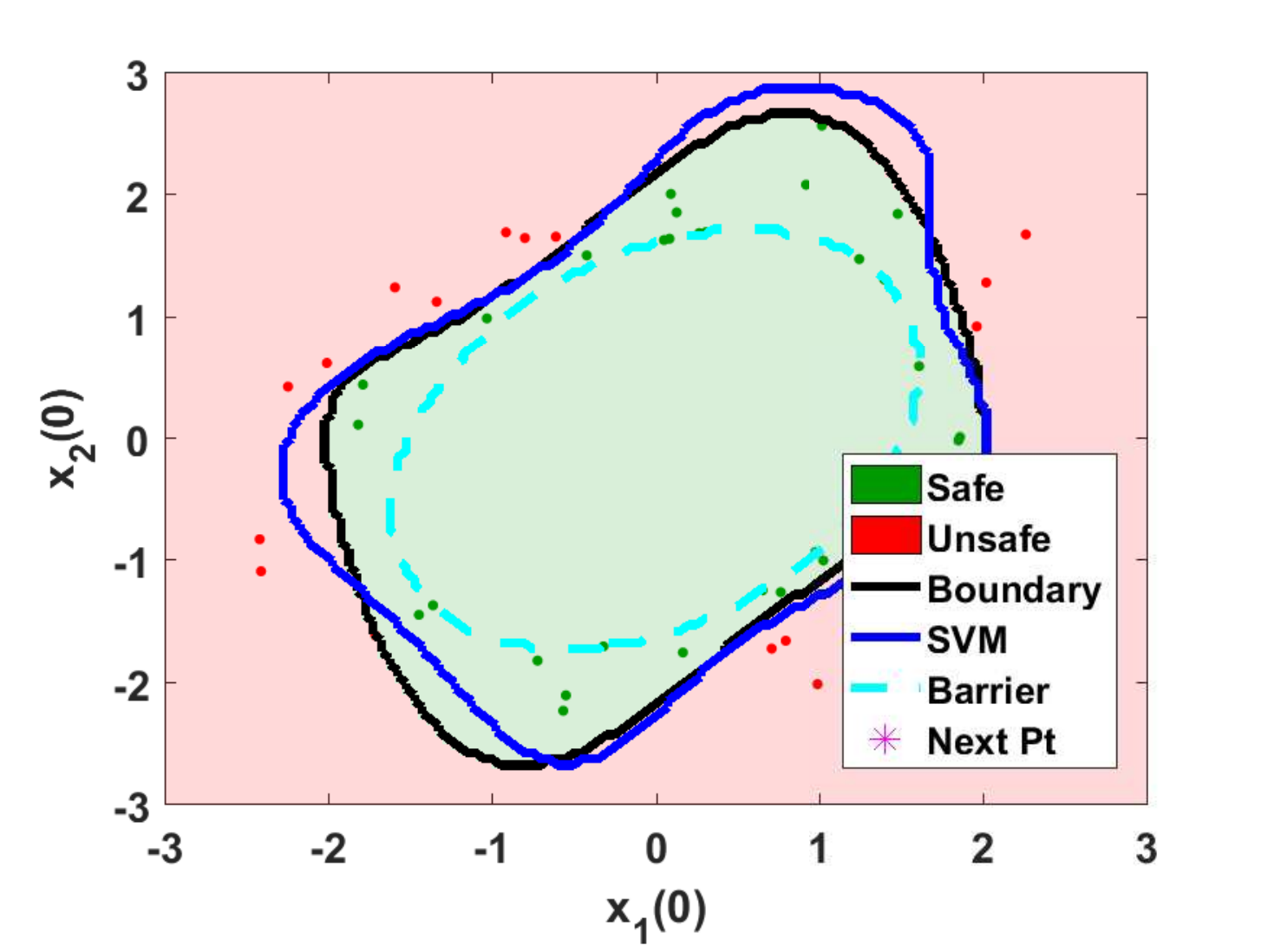}}
	\caption{Initial training set $\mathcal{L}$ and data-driven SVM certificate compared against the analytical certificate produced by the 2nd order Lyapunov function.}
 	\label{f:ex1_init}
\vspace{-0.1in}
\end{figure}

This problem has been well-studied in analytical verification work and suitable $V(x)$ and $\beta$ terms are readily available\cite{Topcu08_PhD,Topcu08_Automatica}.  In order to demonstrate the ability of the data-driven procedure to rapidly improve upon an analytical certificate produced by a poorly-chosen Lyapunov function, a quadratic Lyapunov function is used as the starting condition.  It was shown in \fig{f:vdp1} that a 6th order Lyapunov function is required to accurately approximate the true boundary of the ROA; therefore, the analytical certificate produced by a quadratic Lyapunov function will fail to approximate the true $\Theta_{safe}$.  The discretized set $\Theta_d$ consists of 16,000 points covering $\theta_1: [-3, 3]$ and $\theta_2: [-3, 3]$.  Even with the maximum level set possible with the quadratic $V(x)$, the analytical certificate incorrectly labels nearly 12\% of $\Theta$ as unsafe because these $\vtheta$ locations fall outside the level set defined by the quadratic function.  This is shown in \fig{f:ex1_init}(a).  

\begin{figure}[!]
	\centering
	\subfigure[Selection of the 1st sample in the batch of 10 points]{\includegraphics[width=0.49\columnwidth]{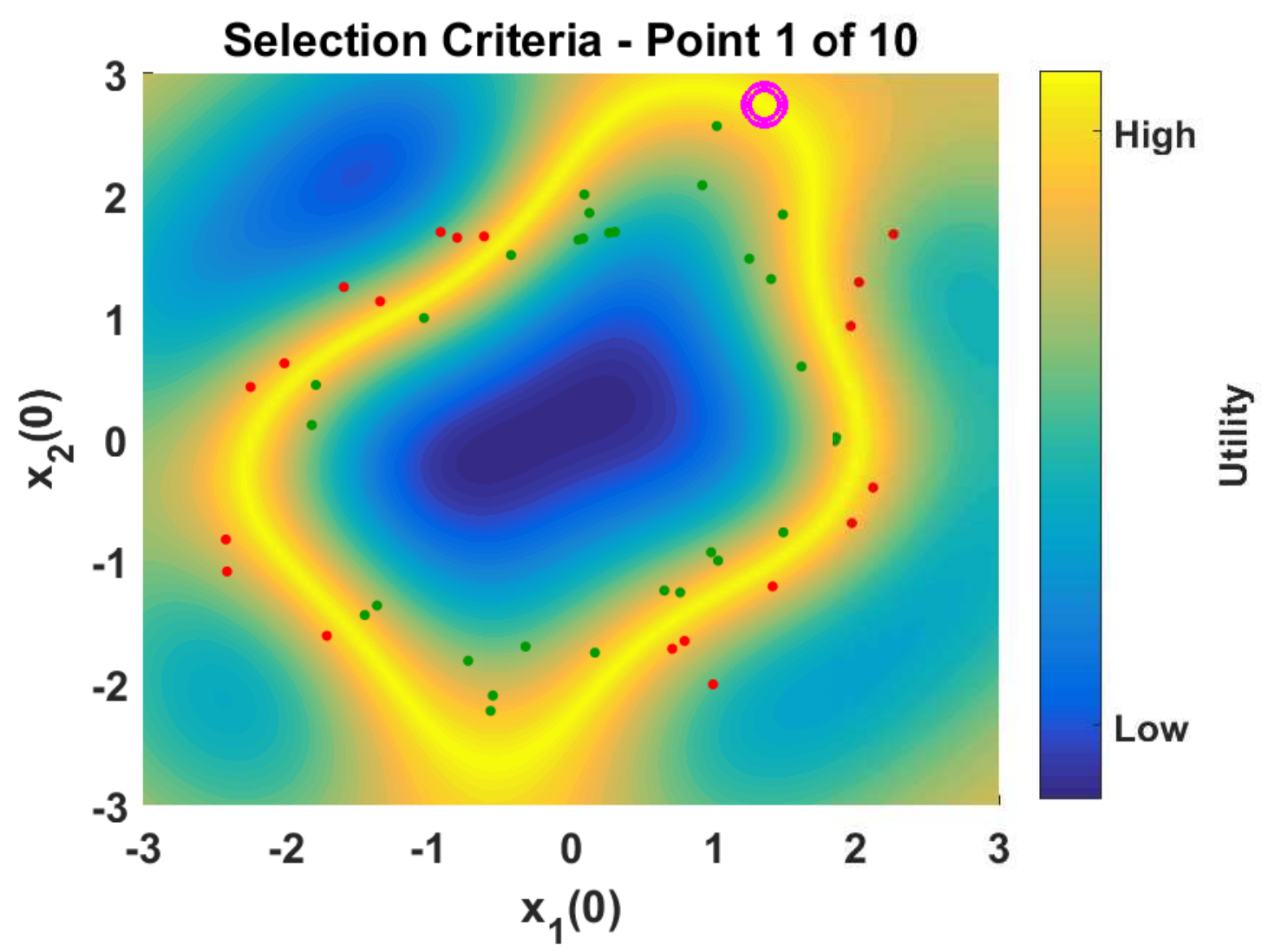}}
	\subfigure[Selection of the 2nd sample]{\includegraphics[width=0.49\columnwidth]{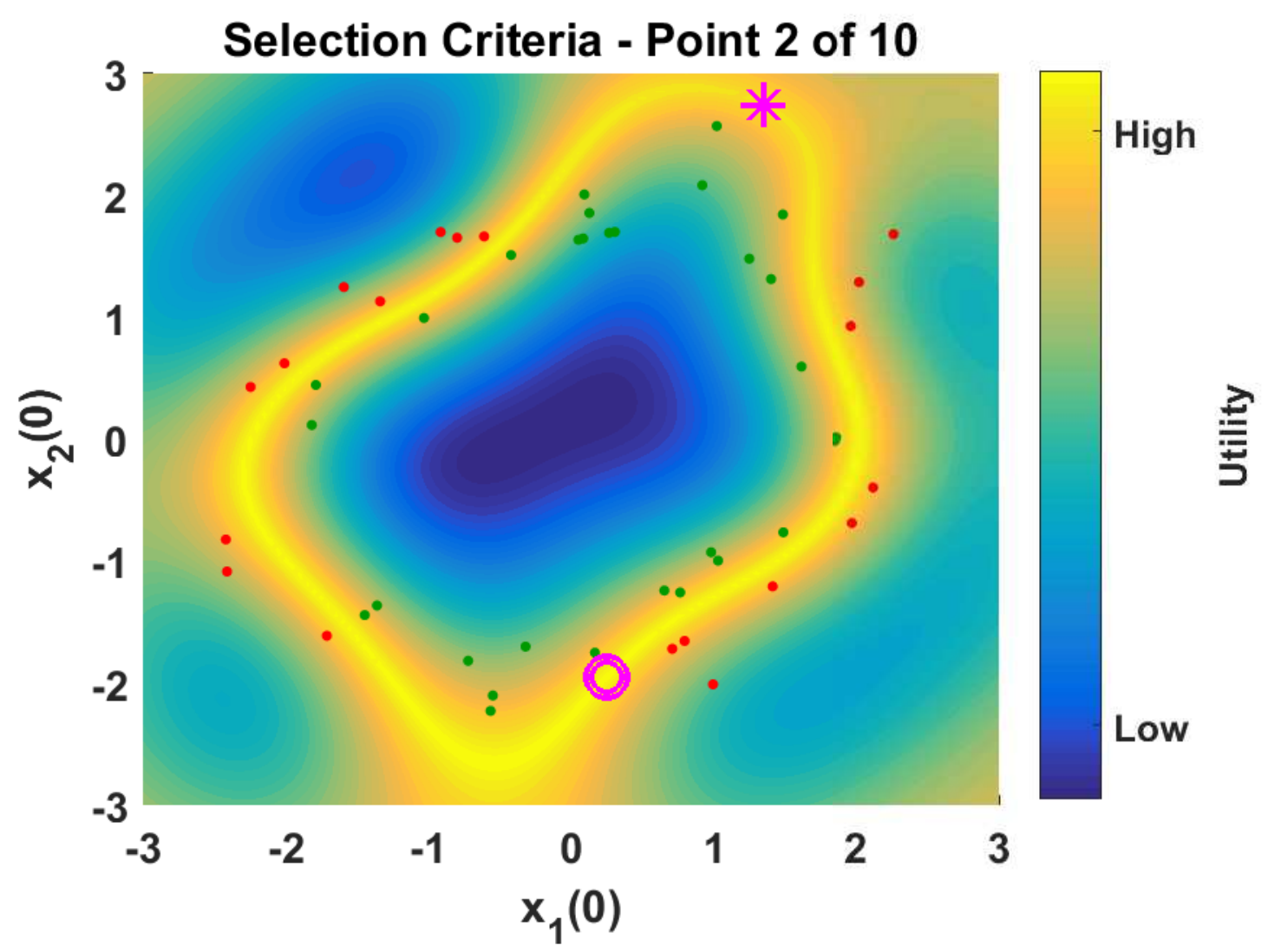}}
	\subfigure[Selection of the 3rd sample]{\includegraphics[width=0.49\columnwidth]{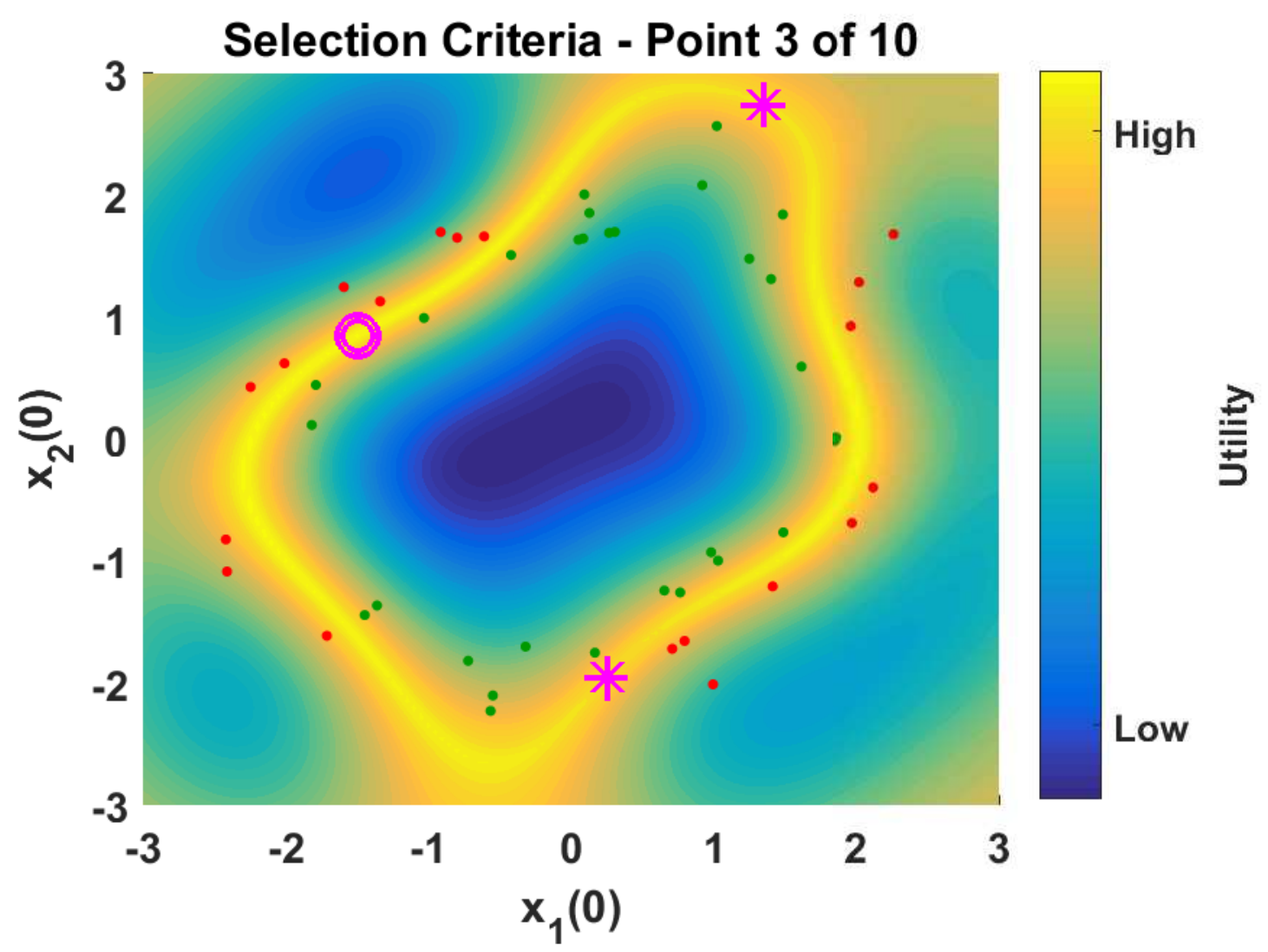}}
	\subfigure[Completed batch of 10 samples]{\includegraphics[width=0.49\columnwidth]{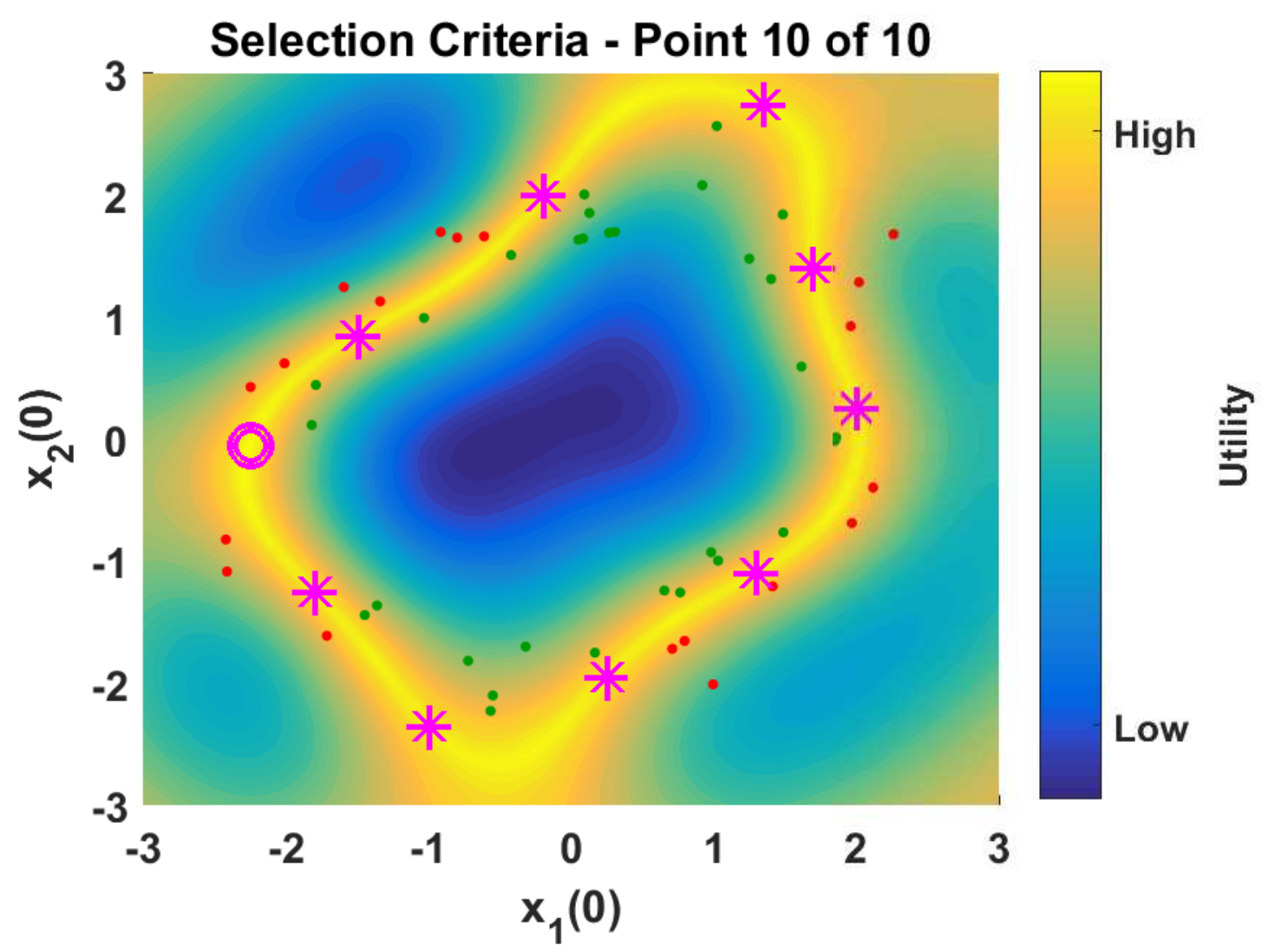}}
	\caption{Selection of the first batch of 10 points according to the expected model change criteria in \cref{eq:batchUnc}.}
 	\label{f:ex1_step1}
\vspace{-0.1in}
\end{figure}
This example considers the batch closed-loop verification procedure from Algorithm \ref{alg:uncSampling}.  The initial SVM model is formed from 50 training samples obtained from the SimLFG sampling algorithm used to produce the simulation-guided analytical barrier certificate\cite{Topcu08_Automatica} seen in \fig{f:ex1_init}(a).  This initial SVM model is pictured in \fig{f:ex1_init}(b).  Once this initial model has been constructed, 20 iterations of \cref{alg:uncSampling} are performed in batches of $M = 10$ points.  The selection of the first batch of 10 points is illustrated in \fig{f:ex1_step1}.  The influence of the diversity measure is clearly visible at the conclusion of the process in \fig{f:ex1_step1}(d).  The samples are spread roughly evenly across the areas of high expected model change in order to prevent redundancy in the batch.  

\begin{figure}[!]
	\centering
	\includegraphics[width=0.6\columnwidth]{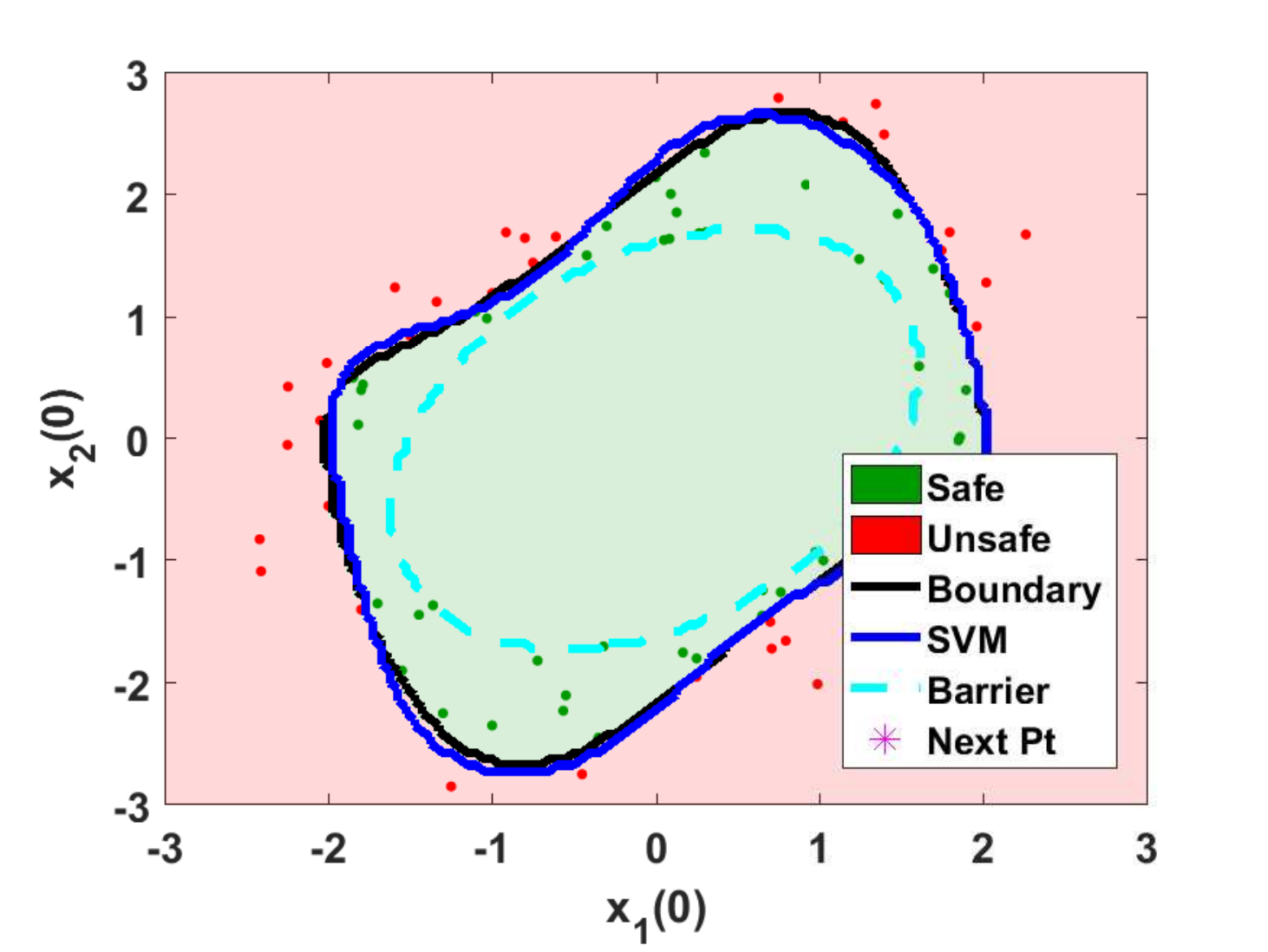}
\vspace{-0.1in}
	\caption{Statistical model of $\Theta_{safe}/\Theta_{fail}$ for the Van der Pol example after only 5 iterations of \cref{alg:uncSampling}.}
	\label{f:ex1_end}
\end{figure}

\begin{figure}[!]
	\centering
	\includegraphics[width=0.6\columnwidth]{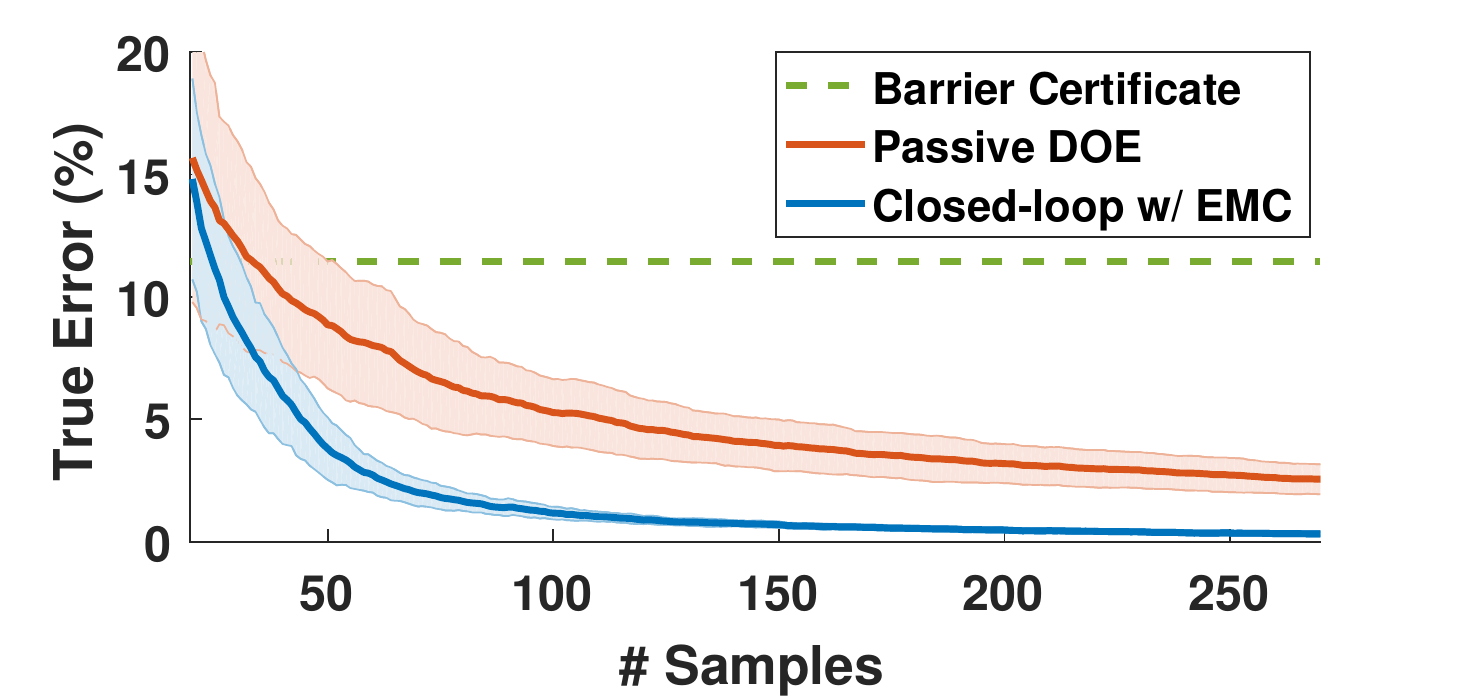}
\vspace{-0.1in}
	\caption{Comparison of the true misclassification error using both active and passive, randomized sampling for the Van der Pol example.  Note that both procedures outperform the analytical certificate, which has a misclassification error of nearly 12\%.}
	\label{f:ex1_stats}
\end{figure}

Even after only 5 iterations (100 training points in total), the statistical classifier converged to a close approximation of the true shape of the ROA, seen in \fig{f:ex1_end}.  The active learning procedure is also compared against a passive procedure that randomly selects batches of 10 samples (without replacement) from $\mathcal{U}$.  \fig{f:ex1_stats} displays the mean and 1-$\sigma$ distribution of the true misclassification error according to 50 repetitions of both of these procedures.  While both procedures already outperform the 12\% misclassification error of the analytical certificate at the very start before obtaining additional samples, active learning ultimately has both a lower mean classification error and 1-$\sigma$ distribution than passive data-driven verification in subsequent iterations.  A video of the active learning procedure can be viewed onlilne here: \url{https://youtu.be/aXTi99edLoA}.  

\begin{figure}[!]
	\centering
	\subfigure[K-fold cross-validation]{\includegraphics[width=0.49\columnwidth]{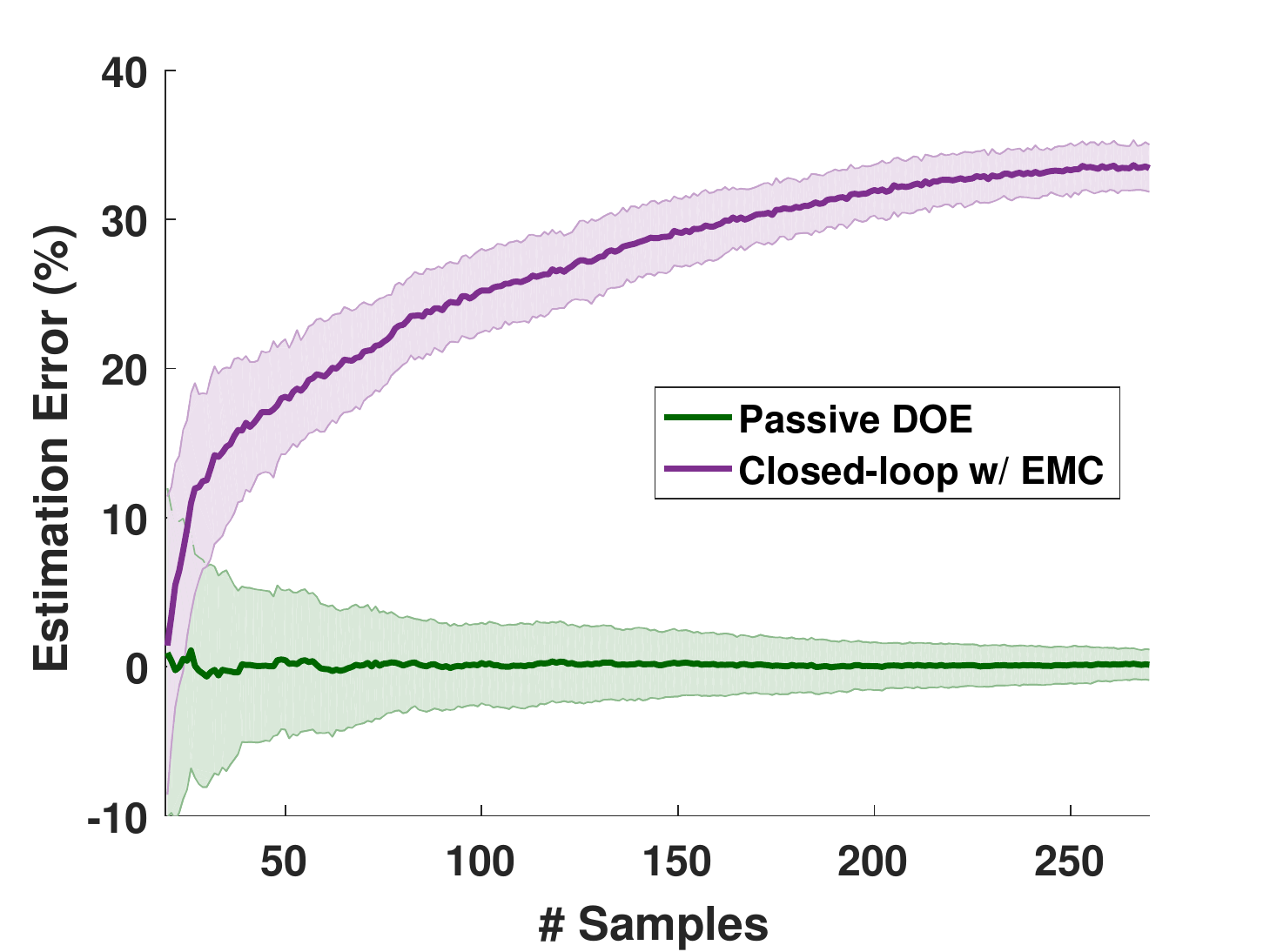}}
	\subfigure[Independent validation set]{\includegraphics[width=0.49\columnwidth]{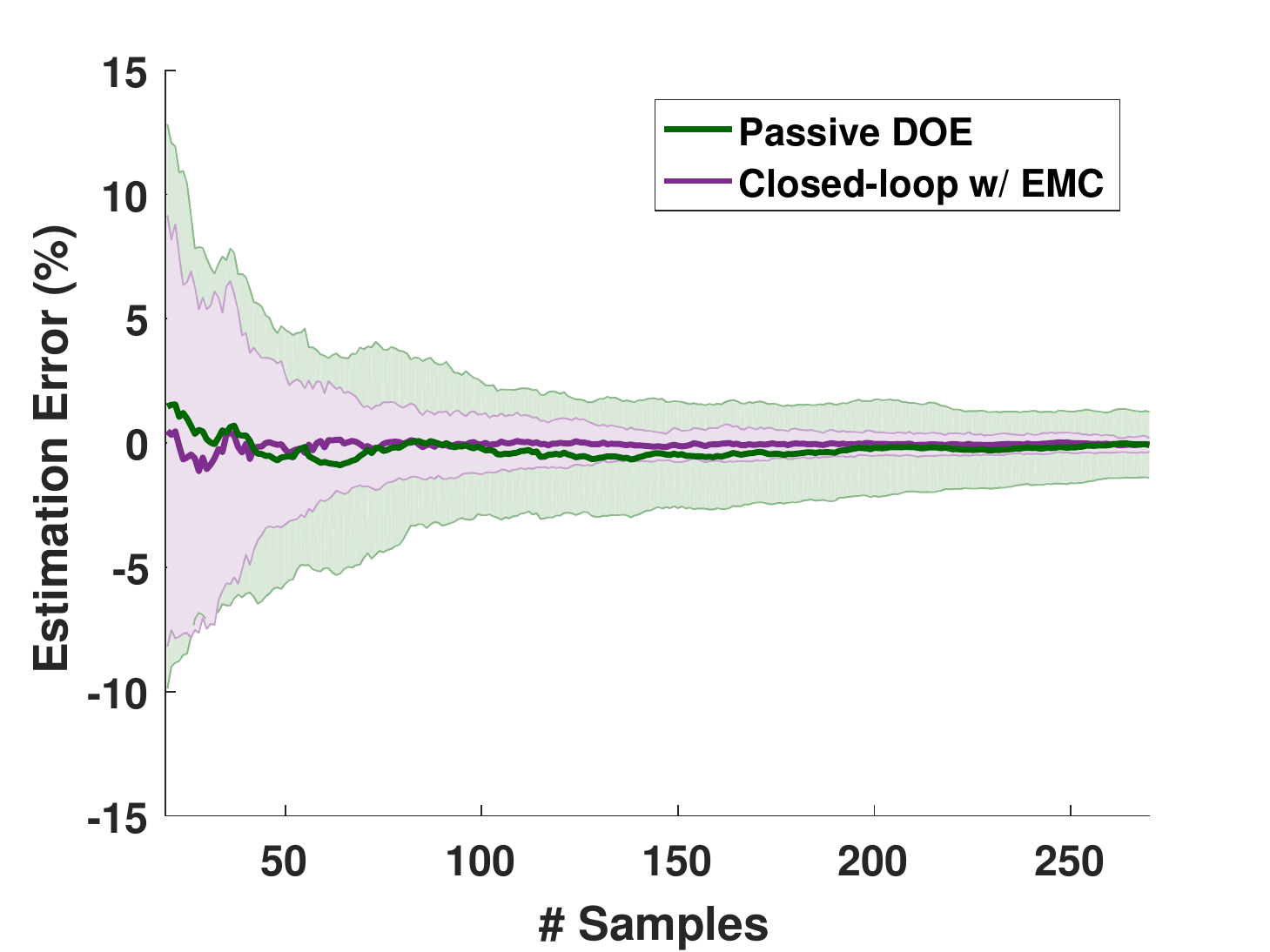}}
	\caption{Comparison of the estimation error produced by k-fold cross-validation and a randomized, independent validation set.}
 	\label{f:ex1_errors}
\vspace{-0.1in}
\end{figure}

Although the results from \fig{f:ex1_stats} quantify the improvement active learning offers over passive approaches, this true misclassification error is unavailable during the actual process as it requires the full set $\Theta_d$ to be known in advance.  Instead, the misclassification error can be approximated using the k-fold cross validation and independent validation set methods described in Section \ref{s:ddverification}-\ref{s:errors}.  The estimation errors produced by these two methods are shown in \fig{f:ex1_errors}.  While k-fold cross validation is more desirable from a computation standpoint as it operates directly on the current training set, the results in \fig{f:ex1_errors}(a) indicate that it is not suitable for use with active learning and closed-loop verification.  The estimated misclassification error increases as the number of samples grows because the closed-loop verification algorithm clusters samples along the approximate boundary.  This clustering will result in the misclassification of a higher proportion of samples than if the data is randomly spread across $\Theta_d$.  In order to obtain an accurate approximation of the misclassification error, an independent validation set must be obtained in parallel to $\mathcal{L}$.

\subsection{Concurrent Learning Adaptive Control System}\label{s:clmrac_ex1}
The second case study is a model reference adaptive control system.  In this problem, a concurrent learning model reference adaptive controller\cite{Chowdhary10_PhD} (CL-MRAC) is used to control an uncertain second order system.  
\begin{align}\label{eq:MRAC}
	\dot{x}_1 & = x_2, \\
	\dot{x}_2 & = (-0.2+\theta_1) x_1 + (-0.2+\theta_2) x_2 + u. \notag
\end{align}
Unlike the previous example, the perturbations $[\theta_1, \theta_2]$ are uncertain parameters in the system.  The adaptive controller estimates these parameters while simultaneously controlling the states to track a desired reference trajectory produced by the linear system in \cref{eq:ref}
\begin{align}\label{eq:ref}
	\dot{x}_{m_1} & = x_{m_2} \\
	\dot{x}_{m_2} & =  -\omega_n^2 x_{m_1} - 2\zeta_n \omega_n x_{m_2} + \omega_n^2 z_{cmd}(t)\ , \notag
\end{align}
with $\zeta_n = 0.5$ and $\omega_n = 1$.  This reference model is excited with reference commands $z_{cmd} = 1$ between 0 and 2 seconds, $z_{cmd} = 1.5$ between 10 and 12 seconds, and $z_{cmd} = -1.5$ between 20 and 22 seconds.  At all other times in the 40 second trajectory length ($T_f = 40$), the reference command defaults to 0.  Note that this reference signal is not persistently exciting.

The control input $u(t)$ consists of three components: the reference input $u_{rm}$, the feedback input $u_{pd}$, and the adaptive input $u_{ad}$.
\begin{equation}\label{eq:inputs}
	u(t) = u_{rm}(t) + u_{pd}(t) - u_{ad}(t)
\end{equation}
In the absence of uncertainties, the reference input and a linear feedback controller are suitable for ensuring closed-loop tracking.
\begin{align}
	u_{rm} & =  -\omega_n^2 x_{m_1} - 2\zeta_n \omega_n x_{m_2} + \omega_n^2 z_{cmd}(t)\\
	u_{pd} & = K_p e_1(t) + K_d e_2(t)
\end{align}
The tracking error $\vec{e}(t) = \vec{x}_m(t) - \vec{x}(t)$ is the difference between the desired reference trajectory and the actual state trajectory.  In this example, the feedback controller gains are set to $K_p = 1.5$ and $K_d = 1.3$.  Due to the presence of the uncertainties, the third $u_{ad}(t)$ component is necessary to ensure the closed-loop system correctly tracks the desired reference trajectory.
\begin{equation}
	u_{ad}(t) = \begin{bmatrix} \hat{\theta}_1(t)  & \hat{\theta}_2(t) \end{bmatrix} \begin{bmatrix} x_1(t) \\ x_2(t) \end{bmatrix} 
\end{equation}
The terms $\hat{\theta}^T = [\hat{\theta}_1(t) \ \ \hat{\theta}_2(t)]$ are the estimated parameters that are updated online according to the concurrent learning adaptive law \cite{Chowdhary10_PhD}
\begin{equation}\label{eq:clmrac}
	\dot{\hat{\theta}} = - \Gamma \vec{x}(t) \vec{e}(t)^T P B - \Gamma_c \sum_{k=1}^{p_{max}} \vec{x}_k \vec{x}_k^T \tilde{\theta}
\end{equation}
with $B^T = [0 \ \ 1]$, $\Gamma = 2$, $\Gamma_c = 0.2$, and $p_{max} = 20$.  The symmetric positive-definite matrix $P$ is obtained from the Lyapunov equation $A^TP + PA = -I$ where $A$ is the nominal open-loop plant (with $\vtheta = 0$).
This adaptive law ensures that both the tracking error $\vec{e}(t)$ and the parameter estimation error $\tilde{\theta}(t) = \hat{\theta}(t) - \theta$ will asymptotically converge to 0 without requiring $z_{cmd}(t)$ to be persistently exciting.  The cornerstone of this convergence is the history stack in the second half of \cref{eq:clmrac} and its periodic update according to the singular value maximizing algorithm\cite{Chowdhary10_PhD}.  The stored datapoints $\vec{x}_k$ are specifically chosen from the trajectory to guarantee $\sum_{k=1}^{p_{max}} \vec{x}_k \vec{x}_k^T > 0$ and are updated only if the new datapoint will improve the rate of convergence.  While this adaptive control law is used to improve the closed-loop response of the system, it transforms the second-order linear plant into a nonlinear system.  The history stack in particular greatly complicates analysis of the closed-loop response due its periodic, but non-uniform updates of the saved datapoints.

Although the CL-MRAC controller ensures asymptotic stability of the closed-loop system, safety of the system in this example is measured by whether the state $x_1(t)$ remains within unit error of the reference $x_{m_1}(t)$.  Described in metric temporal logic format, this performance requirement states 
\begin{equation}
	\varphi_{bound} = \Box_{[0, T_{f}]} \ (1 - |e_1[t]| \geq 0)\ .
\end{equation}
Even though the system is guaranteed to the stable with respect to the tracking error $e(t)$, this stability does not directly guarantee $e_1(t)$ will remain within the specified bounds.  The goal of the verification process is to identify which $\vtheta$ vectors will not cause the trajectory to exceed those bounds and which will.  Similar to before, an analytical barrier certificate was previously obtained in prior work\cite{Quindlen16_ACC} and is based upon the quadratic Lyapunov function commonly used for adaptive systems.  Unlike the first example, there are no alternative Lyapunov functions available; a general higher-order Lyapunov function for CL-MRAC systems has not been found.  

\begin{figure}[!]
	\centering
	\includegraphics[width=0.7\columnwidth]{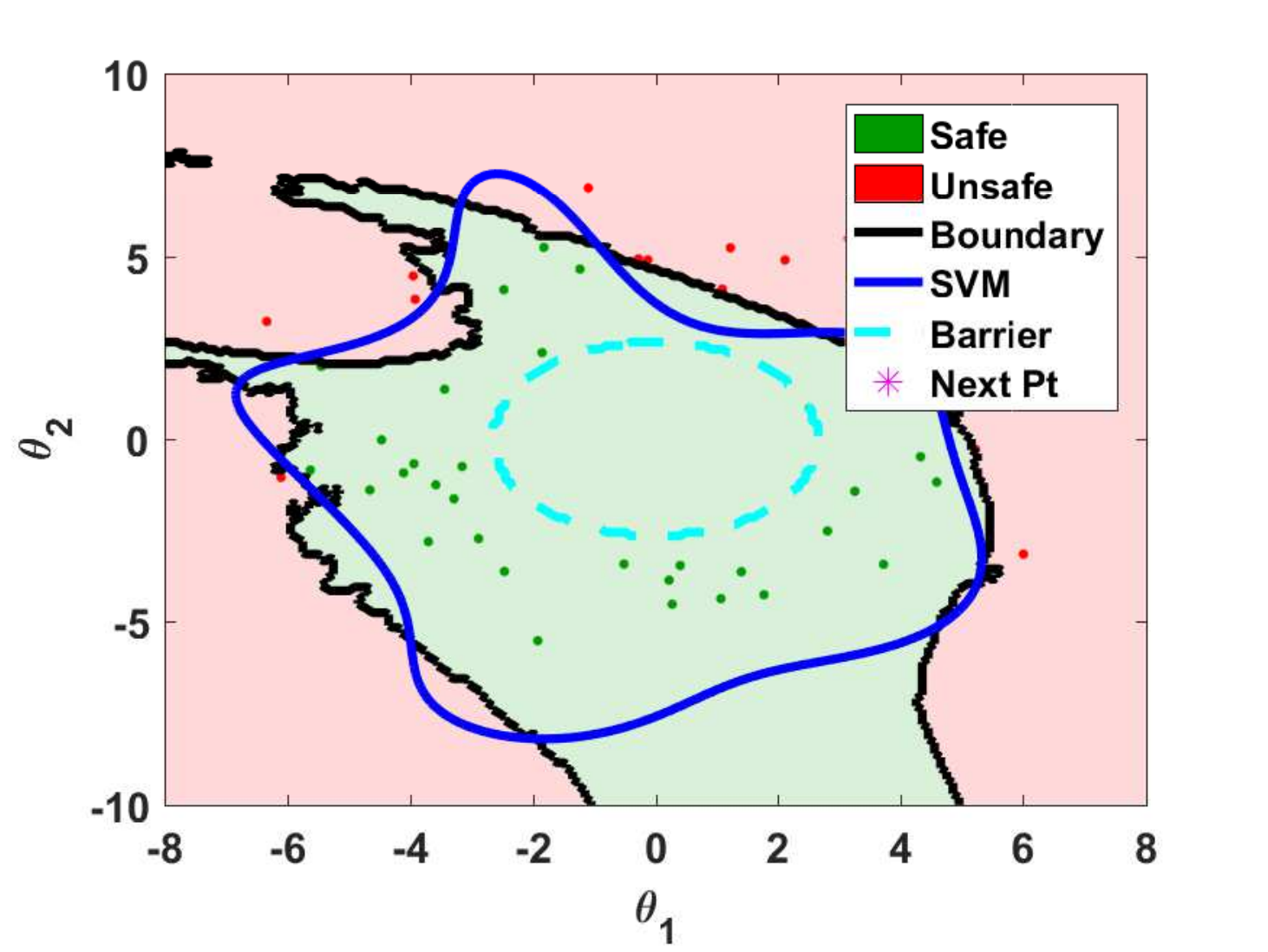}
	\vspace{-0.1in}
	\caption{The initial training set $\mathcal{L}$ and SVM for the system in \cref{eq:MRAC}.}
	\label{f:ex2_init}
	%
	\centering
	\includegraphics[width=0.7\columnwidth]{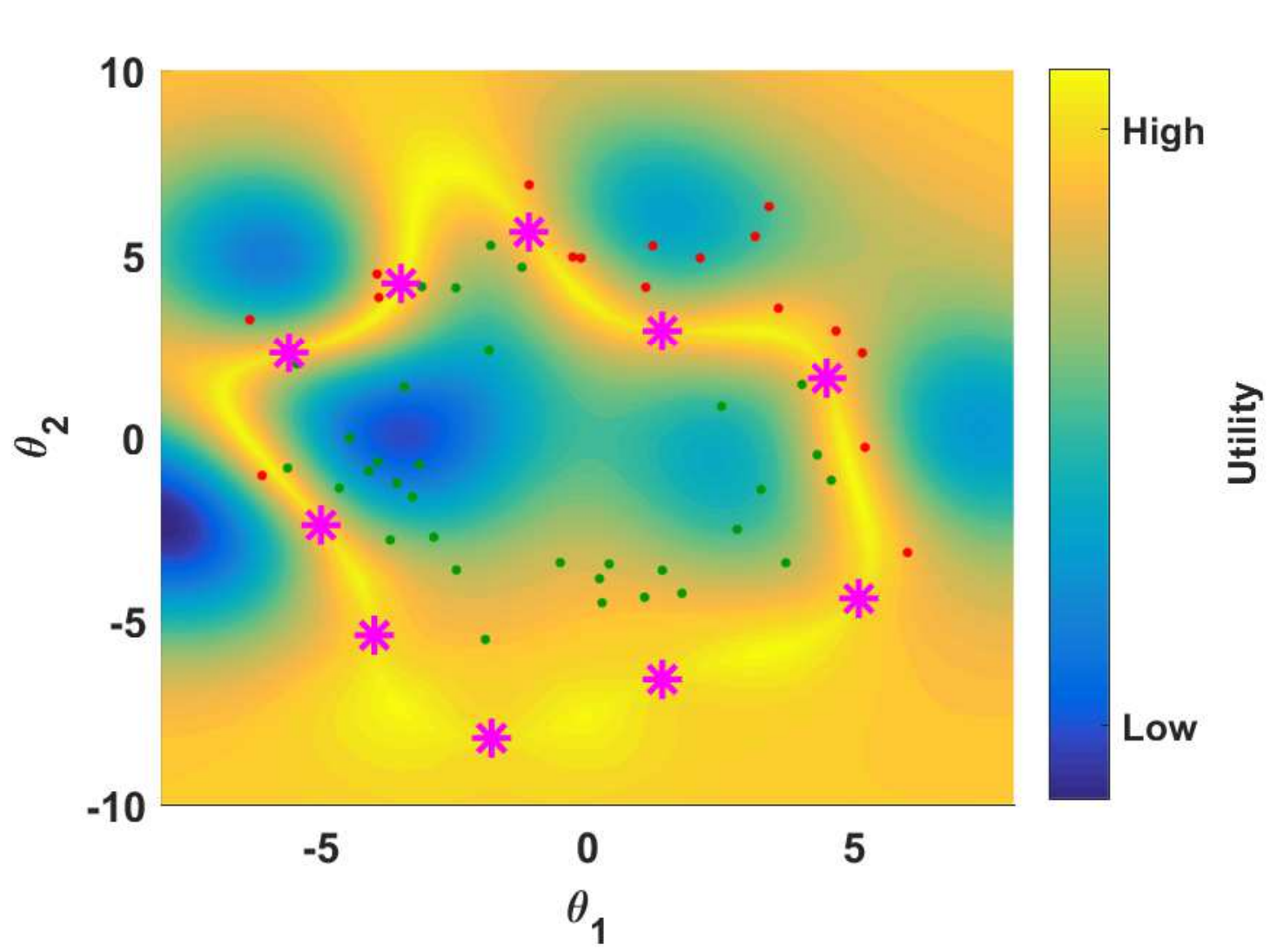}
	\vspace{-0.1in}
	\caption{Selection criteria for the first batch of 10 samples using \cref{alg:uncSampling}.}
	\label{f:ex2_sample}
\end{figure}

\begin{figure}[!]
	\centering
	\includegraphics[width=0.75\columnwidth]{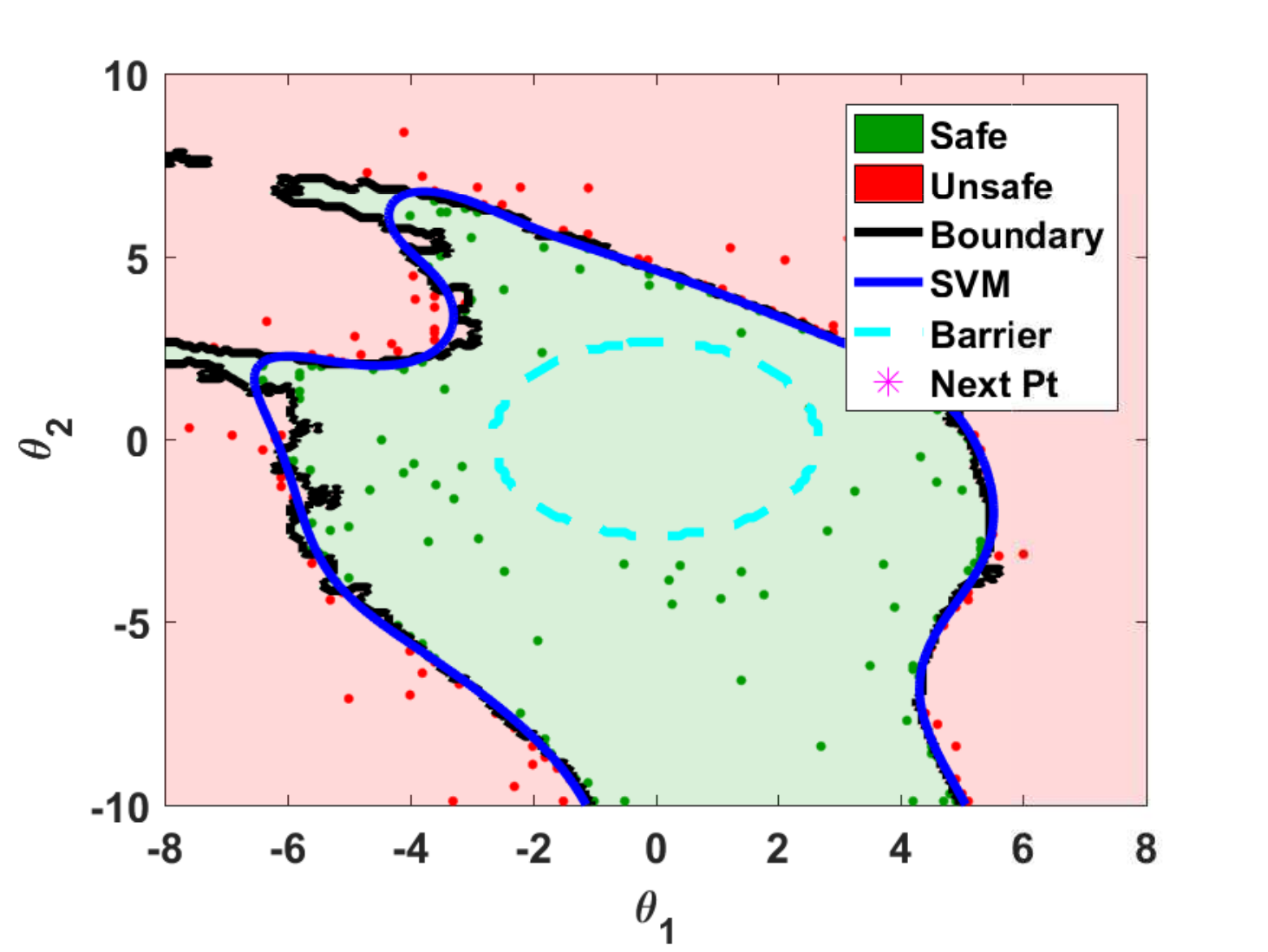}
	\vspace{-0.1in}
	\caption{Statistical certificate after 20 iterations of \cref{alg:uncSampling}.}
	\label{f:ex2_end}

	\centering
	\includegraphics[width=0.7\columnwidth]{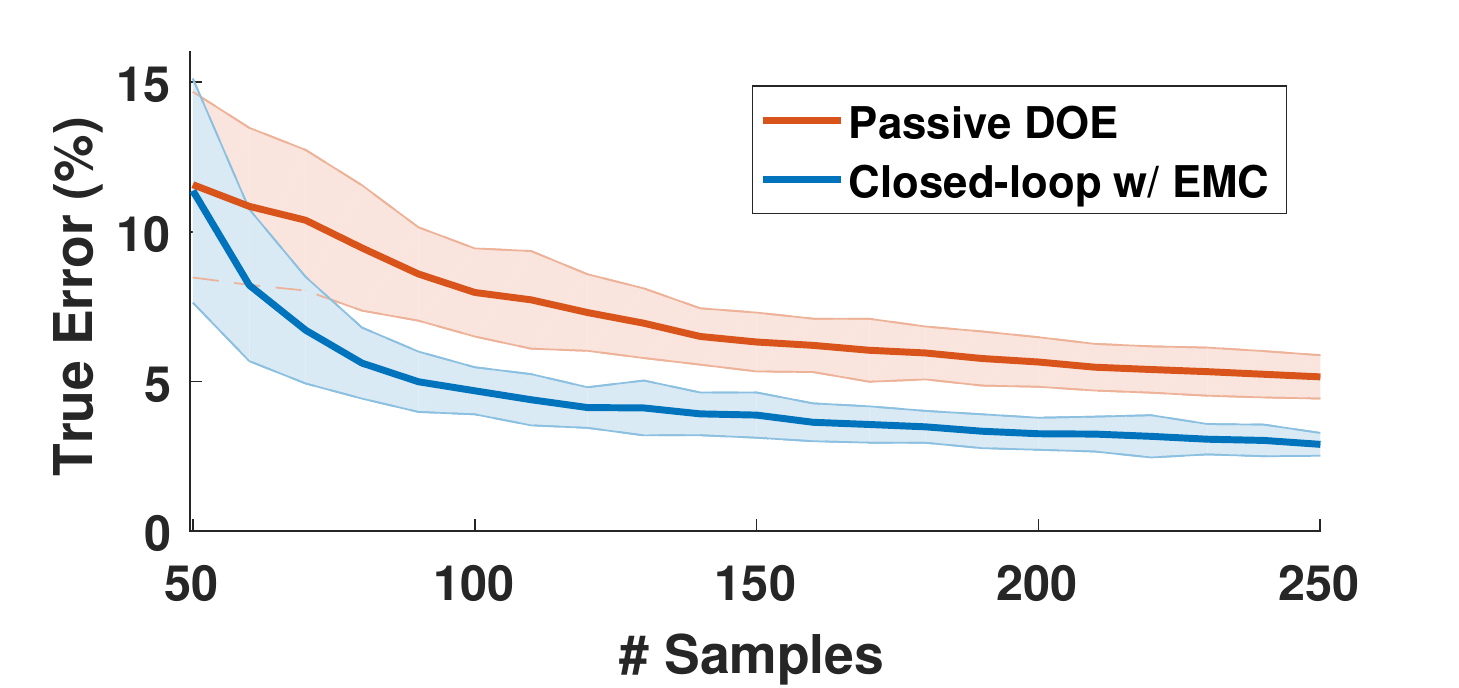}
	\vspace{-0.1in}
	\caption{Comparison of the true misclassification error for the system in (\ref{eq:MRAC}) using both active and passive, randomized sampling. Both of these outperform the analytical certificate, which misclassifies 35\% of $\Theta_{safe}$ as ``unsafe''.}
	\label{f:ex2_stats}
\end{figure}

Twenty iterations of the closed-loop data-driven procedure are performed in batches of 10 points.  In this problem, the unlabeled set $\mathcal{U}$ consists of 36,400 points covering $\theta_1: [-8,8]$ and $\theta_2: [-10,10]$.  The initial training set $\mathcal{L}$ consists of 50 randomly-selected points and is used to form the initial SVM, pictured in \fig{f:ex2_init}.  The conservativeness of the 2nd order Lyapunov function-based barrier certificate is readily apparent in that figure.  \fig{f:ex2_sample} overlays the selection criteria from \cref{eq:batchUnc} over $\Theta$ and displays the 10 points that make up the first batch.  As before, the selected sample locations are spread across regions of high expected model change.  \fig{f:ex2_end} shows the data-driven classifier after 20 iterations.  While the nonconvex shape of the $\Theta_{safe}/\Theta_{fail}$ boundary is significantly more challenging than the shape of the Van der Pol oscillator's limit cycle, the statistical model still converges toward the true boundary and minimizes the prediction error.  The mean and 1-$\sigma$ distribution of the true misclassification error for 50 repetitions are shown in \fig{f:ex2_stats} and reinforce this observation.  A video of the active learning procedure can be viewed online here:\\ \url{https://youtu.be/ID_fZ2SKpIk}.


\begin{figure}[!]
	\centering
	\includegraphics[width=0.7\columnwidth]{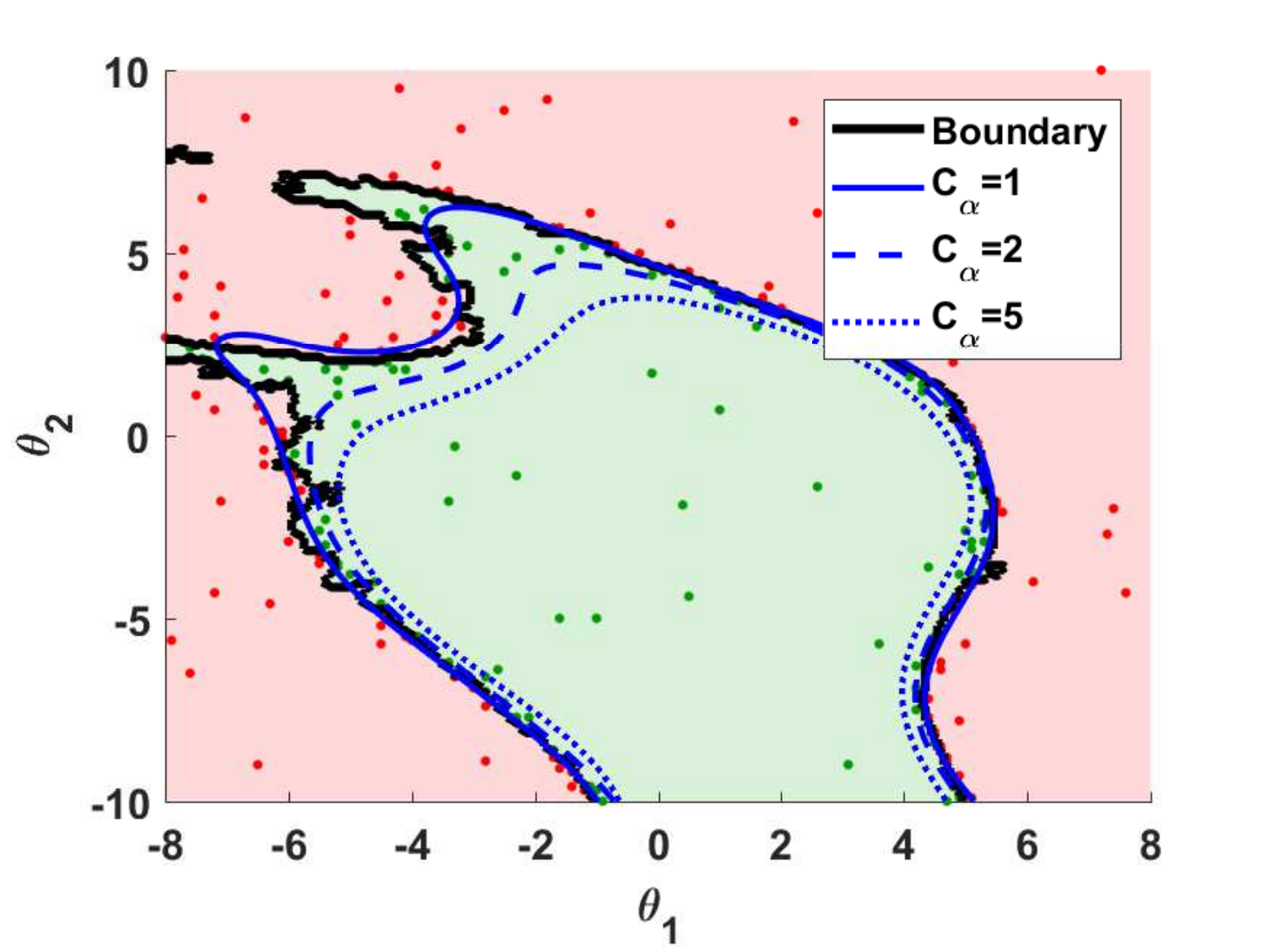}
	\vspace{-0.1in}
	\caption{Comparison of the data-driven classifier given different penalties on unsafe misclassification errors.  The terms $C_{\alpha}=1,2,5$ refer to the ratio of training penalty placed on mislabeling unsafe locations as ``safe'' against mislabeling safe locations as ``unsafe''.}
	\label{f:ex2_costs}
\end{figure}

Although active learning is shown to reduce the true misclassification error over analytical certificates, there does exist the possibility of unsafe misclassification errors - $\vtheta$ locations that actually fail to meet the requirements accidentally labeled as ``safe.''  In the training process described in Section \ref{s:ddverification}, the $C_{\alpha}$ weighting term from \cref{eq:training2} can be adjusted so that these unsafe errors are penalized higher than the reverse.  This will introduce conservativeness into the construction of the classifier, seen in \fig{f:ex2_costs}.  While this does increase the true misclassification error from 2.58\% at $C_{\alpha}=1$ to 9.50\% at $C_{\alpha}=5$, it reduces the percentage of unsafe misclassification errors from 1.30\% to 0.  This also holds true when the results are compared on an independent validation set of equal size of $\mathcal{L}$ - estimated total error increases from 4.44\% to 9.60\%, but unsafe error decreases from 2.80\% to 0.  The choice of $C_{\alpha}$ is left to the designer, but $C_{\alpha} \neq 1$ should only be used after completion of the active sampling procedure to avoid handicapping the search for the true boundary.

\subsection{Adaptive Control System with Control Saturation}
The third example is a variant of the same CL-MRAC system from the preceding example with new performance requirements and more complex constraints.  The open-loop system plant is the same as in \cref{eq:MRAC}; however, the control input $u(t)$ is saturated with limits $-u_{max} \leq u(t) \leq u_{max}$.  Likewise, the perturbations $\vtheta$ are different in this new problem.  In addition to the same parametric uncertainties $[\theta_1, \theta_2]$, there is also uncertainty in the initial position $x_1(0)$.  Various levels of possible control saturation limits $u_{max}$ are also considered during the verification process as a 4th dimension of $\vtheta$, although $u_{max}$ would generally be known at run-time.  

The same CL-MRAC control law from \cref{eq:inputs} and \cref{eq:clmrac} is used, but the presence of the control saturation can lead to instability in the adaptation of $\hat{\theta}$ if left unaddressed.  In order to counter the adverse effects of control saturation on the adaptation, pseudo-control hedging (PCH)\cite{Kannan10_CDC} is used to augment the adaptive system.  Pseudo-control hedging creates a hedge input $\nu_h$ from the difference between the desired control input before saturation $u_{des}(t)$ and the control saturation limit $u_{max}$.  This desired control input before saturation $u_{des}(t)$ is produced by the same equation \cref{eq:inputs} while the actual control input after saturation is $u(t)$.
\begin{equation}\label{eq:inputs2}
	u_{des}(t) = u_{rm}(t) + u_{pd}(t) - u_{ad}(t)
\end{equation}
\begin{equation}
	\nu_h = \begin{cases} u_{max} - u_{des}(t) & \text{ if } u_{des}(t) > u_{max} \\ 0 & \text{ otherwise } \\ -u_{max} - u_{des}(t) & \text{ if } u_{des}(t) < -u_{max} \end{cases}
\end{equation}
\begin{equation}
	u(t) = \begin{cases} u_{max} & \text{ if } u_{des}(t) > u_{max} \\ u_{des}(t) & \text{ otherwise } \\ -u_{max} & \text{ if } u_{des}(t) < -u_{max} \end{cases}
\end{equation}
The issue that arises due to saturation is the controller can no longer perfectly track the reference trajectory $\vec{x}_m(t)$ when the control input $u(t)$ is saturated.  Instead, the pseudo-control hedge $\nu_h$ modifies the reference model in order to prevent the saturation from negatively affecting the tracking error $\vec{e}(t) = \vec{x}_m(t) - \vec{x}(t)$.  In this example, the PCH-modified reference model becomes
\begin{align}\label{eq:ref2}
	\dot{x}_{m_1} & = x_{m_2} \\
	\dot{x}_{m_2} & =  -\omega_n^2 x_{m_1} - 2\zeta_n \omega_n x_{m_2} + \omega_n^2 z_{cmd}(t) - \nu_h \ . \notag
\end{align}

In addition to the changes to the plant and control architecture, the closed-loop system is also expected to satisfy a new set of temporal requirements.  More specifically, the system's trajectory must satisfy three conditions in order for it to be labeled as ``safe''.  First, the trajectory must fall within the interval $x_1(t) \in [0.7, 1.3]$ at some point between $t = 2$ and $t=3$ seconds.
\begin{equation}
	\varphi_1 = \Diamond_{[2,3]} \ (x_1[t] - 0.7 \geq 0) \wedge \Diamond_{[2,3]} \ (1.3 - x_1[t] \geq 0)
\end{equation}
Next, the trajectory must satisfy a similar requirement between $t=12$ and $t=13$ seconds, where the trajectory should reach $x_1(t) \in [1.1, 1.7]$ at some point in the time interval.  
\begin{equation}
	\varphi_2 = \Diamond_{[12,13]} \ (x_1[t] - 1.1 \geq 0) \wedge \Diamond_{[2,3]} \ (1.7 - x_1[t] \geq 0)
\end{equation}
Finally, the trajectory should reach some point in the interval $x_1(t) \in [-1.6, -1.2]$ at $t=22.5$ seconds.  This requirement is modeled as 
\begin{equation}
	\varphi_3 = \Box_{[22.4,22.6]} \ (x_1[t] + 1.6 \geq 0) \wedge \Box_{[22.4,22.6]} \ (-1.2 - x_1[t] \geq 0)\ .
\end{equation}
The trajectory is only labeled as ``safe'' if it is able to satisfy all three of the requirements.
\begin{equation}
	\varphi = \varphi_1 \wedge \varphi_2 \wedge \varphi_3
\end{equation}
Although the nominal reference model was designed to meet all three temporal requirements, PCH adjustments to the reference model will change the resulting reference trajectory and can accidentally lead both the true and reference trajectories to fail to meet one or all of the temporal requirements.  Therefore, it is critically important to examine whether the actual trajectory $\vec{x}(t)$ satisfies those requirements and not simply focus on the tracking error $\vec{e}(t)$ as in the second example problem.

\begin{figure}
	\centering
	\includegraphics[width=0.7\columnwidth]{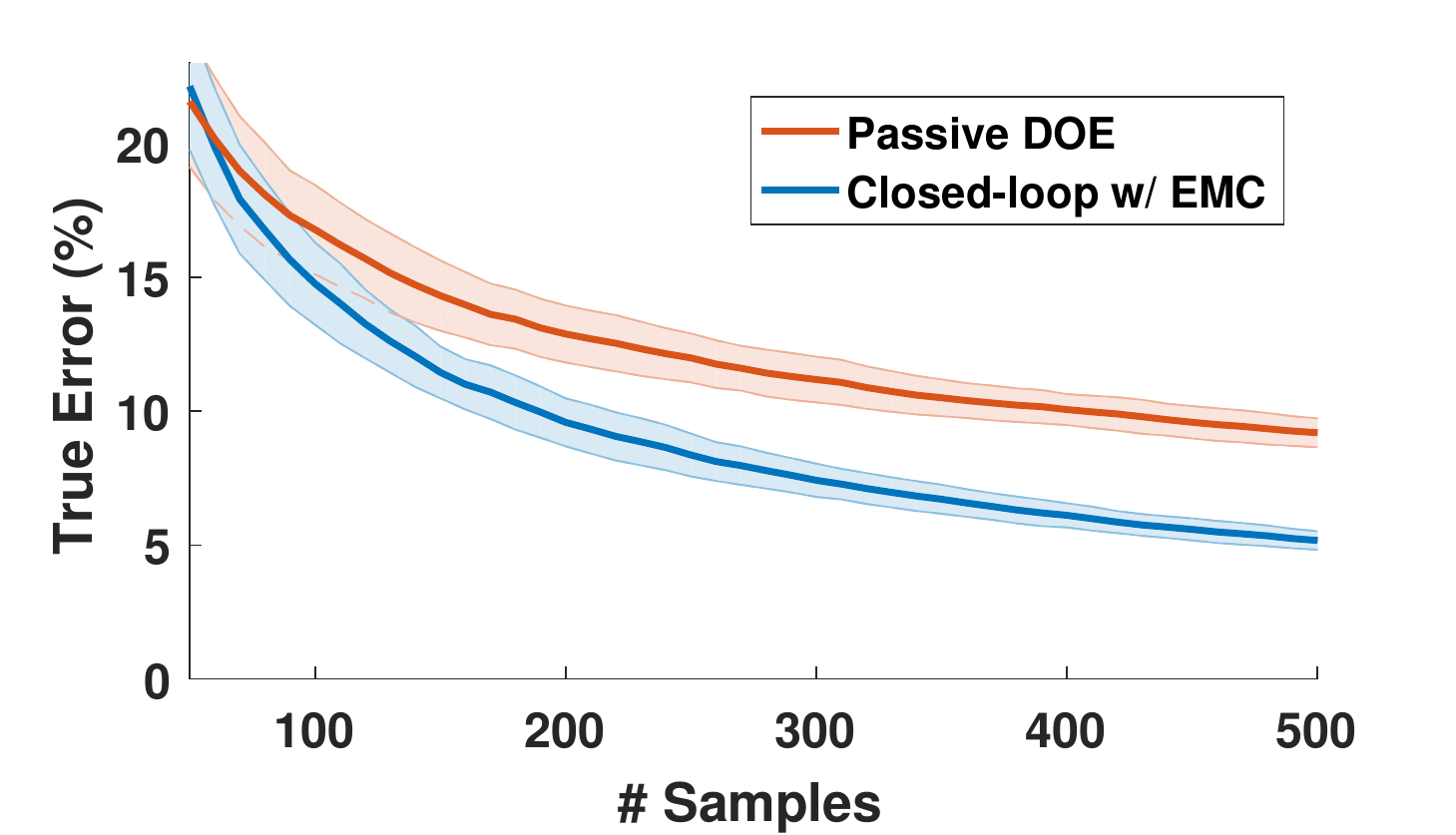}
	\vspace{-0.1in}
	\caption{Comparison of the prediction error for the constrained adaptive control system example using both active and passive, randomized sampling procedures.  Note that error bars are only shown for even-numbered iterations for ease of viewing.  Unlike the previous two examples, an analytical barrier certificate was not found and cannot be compared against the results.}
	\label{f:ex3_stats}
\end{figure}

Just as in the previous two case studies, active data-driven verification learns a statistical certificate of the satisfaction of performance requirements subject to a bound on the number of simulations.  Unlike the previous two examples, an analytical barrier certificate cannot be computed due to the complexities of the closed-loop system and performance requirements.  This highlights that data-driven verification can be applied to a wider class of possible certification requirements than is possible with the existing barrier certificate techniques.  The procedure constructs a 4-dimensional discretized grid $\Theta_d$ of 1.285 million possible sample locations, with each dimension corresponding to $\theta_1: [-5,5]$, $\theta_2:[-5,5]$, $x_1(0): [-1,1]$ and $u_{max}: [3,8]$.  From this grid, the active sampling procedure obtains 50 initial training samples and performs 45 iterations in batches of 10 points.  At the conclusion of the process, the statistical certificate is constructed using only 500 training samples and predicts performance satisfaction over the remaining set of unobserved values.  The mean and 1-$\sigma$ prediction error averaged over 100 randomly-initialized repetitions is shown in \fig{f:ex3_stats}.  Just as with the previous case studies, active learning is able to outperform passive data-driven verification with both a lower mean and standard deviation from that mean as the number of iterations grows.  At the conclusion of the process, the active learning procedures produces an average true misclassification error of 5.17\% and estimated validation error of 5.15\% compared against 9.16\%/9.11\% for the passive procedure.  Even for this large 4D system, active data-driven verification is shown to be a suitable tool for examining the performance of a complex adaptive system.

\section{Conclusion}
This work has presented a data-driven approach for verification of uncertain dynamical systems.  The goal of the procedure is to estimate the boundary that separates safe perturbations from unsafe ones.  The core of the data-driven verification procedure is a support vector machine (SVM) model to classify all the points in $\Theta$ and an active sampling procedure to iteratively improve upon it.  Section \ref{s:ddverification} described the construction of the statistical certificate and the importance of the training data.  Meanwhile, closed-loop algorithms for active selection of future training points were presented in Section \ref{s:active}.  These closed-loop sampling processes select sample points that will maximize the expected change in the model in order to minimize the misclassification error.  The utility of the closed-loop data-driven verification procedures was successfully demonstrated on multiple case studies of nonlinear and adaptive control systems.  When compared against passive procedures that rely upon random sampling, active learning results in both lower a mean and distribution of the prediction errors.  

Data-driven verification is a supplementary approach for analytical verification of dynamical systems.  In many cases, it is either not obvious or possible to obtain the analytical certificate that identifies the full boundary, resulting in overly-conservative approximations.  In these cases, statistical classifiers can be used a direct replacement of these barrier certificates or provide feedback for subsequent corrections.  Additionally, data-driven verification can be applied to a larger class of problems than possible with analytical techniques.  In some applications, an equivalent analytical certificate does not exist because there are no suitable analytical functions to provably bound the response.  

While the lack of a governing analytical function allows data-driven methods to apply to a wider range of possible systems, it is important to recognize that it comes at a cost.  Unlike analytically-verified certificates, data-driven certificates can accidentally mislabel unsafe perturbations as ``safe''; therefore, safe predictions from a data-driven certificate are not as strong and may be incorrect.  The desire to minimize the likelihood of these misclassifications was one of the primary motivations for active learning.  Ongoing research has focused on additional sampling methods and retraining steps to combine with the active learning procedure in this work.  Ultimately, this paper presents a foundation for more advanced data-driven verification methods to certify the safety of complex dynamical systems subject to various sources of uncertainty.

\balance
\bibliographystyle{aiaa}
\bibliography{BIB_Jack/JQ_15,BIB_all/ACL_Publications}

\end{document}